# A Long-term Study of Mrk 50: Appearance and Disappearance of Soft Excess

Narendranath Layek[1,2,5], Prantik Nandi[3], Sachindra Naik[1], and Arghajit Jana[4]
[1] Astronomy and Astrophysics Division, Physical Research Laboratory, Navrangpura, Ahmedabad 380009, Gujarat, India; narendral@prl.res.in, narendranathlayek2017@gmail.com
[2] Indian Institute of Technology Gandhinagar, Palaj, Gandhinagar 382055, Gujarat, India
[3] Indian Centre for Space Physics, 466 Barakhola, Netaji Nagar, Kolkata, 700099, India
[4] Instituto de Estudios Astrofísicos, Facultad de Ingeniería y Ciencias, Universidad Diego Portales, Av. Ejército Libertador 441, Santiago, Chile

## Abstract

We present an extensive temporal and spectral study of the Seyfert 1 AGN Mrk 50 using 15 yr (2007–2022) of multiwavelength observations from XMM-Newton, Swift, and NuSTAR for the first time. From the timing analysis, we found that the source exhibited variability of ∼20% during the 2007 observation, which reduced to below 10% in the subsequent observations and became nonvariable in the observations from 2010 onward. From the spectral study, we found that the spectra are nearly featureless. Nondetection of absorption in the low-energy domain during the 15 yr of observation infers the absence of obscuration around the central engine, rendering the nucleus a "bare" type. A prominent soft X-ray excess below 2 keV was detected in the source spectrum during the observations between 2007 and 2010, which vanished during the later observations. To describe the nature of the soft excess, we use two physical models, such as warm Comptonization and blurred reflection from the ionized accretion disk. Both of the physical models explain the nature and origin of the soft excess in this source. Our analysis found that Mrk 50 accretes at a sub-Eddington accretion rate ($\lambda_{\rm Edd}$ = 0.13–0.02) during all of the observations used in this work.

*Unified Astronomy Thesaurus concepts:* Active galactic nuclei (16); X-ray active galactic nuclei (2035); Seyfert galaxies (1447)

## 1. Introduction

Active galactic nuclei (AGNs) are among the most luminous and energetic sources in the Universe. The extremely high luminosity of the AGNs is understood to arise from the accretion of matter onto supermassive black holes (SMBHs) residing at the center of the host galaxies (M. J. Rees 1984). The SMBH typically has a mass ranging from $10^5$–$10^9$ $M_\odot$ (J. Kormendy & D. Richstone 1995; B. M. Peterson et al. 2004; Event Horizon Telescope Collaboration et al. 2019). The AGNs emit radiation across the entire range of the electromagnetic spectrum, from radio waves to high-energy γ-rays. According to the standard accretion disk theory (N. I. Shakura & R. A. Sunyaev 1973), the thermal emission from the disk is mostly emitted in the ultraviolet (UV) band of the electromagnetic spectrum. Viscous dissipation in the disk generates heat, which is then radiated in the optical/UV regime for black hole (BH) masses typical of AGNs (W.-H. Sun & M. A. Malkan 1989). The X-ray emission from the AGN serves as an important tool for investigating physical processes in extreme gravity, as it is thought to originate from the innermost region of the accretion disk. The X-ray spectrum of AGNs is primarily dominated by a power-law continuum produced through the inverse Compton scattering of the seed optical/UV photons from the accretion disk by a hot ($T \sim 10^9$ K) optically thin corona (N. I. Shakura & R. A. Sunyaev 1973; F. Haardt & L. Maraschi 1993; H. Netzer 2013) located near the black hole (F. Haardt & L. Maraschi 1991; R. Narayan & I. Yi 1994;

S. Chakrabarti & L. G. Titarchuk 1995; C. Done et al. 2007). The power-law continuum often shows a high-energy exponential cutoff, usually around a few hundred kiloelectronvolts (R. A. Sunyaev & L. G. Titarchuk 1980). This feature is directly related to the temperature and optical depth of the plasma of hot electrons responsible for the power-law emission. In addition to the power-law continuum with exponential cutoff, the other spectral features, such as reflection (R. R. Ross & A. C. Fabian 1993, 2005; J. García et al. 2013, 2014), low-energy absorption, and often an excess in the soft X-ray ranges (below 2 keV) (J. P. Halpern 1984; K. A. Arnaud et al. 1985; K. P. Singh et al. 1985; T. J. Turner & K. A. Pounds 1989), are also present in the X-ray spectra of the AGNs. The reflection features consist mainly of two components, such as a Compton hump, visible in the 15–50 keV range with a peak at around ∼30 keV (J. H. Krolik 1999) and an iron $K_\alpha$ emission line at ∼6.4 keV (I. M. George & A. C. Fabian 1991; G. Matt et al. 1991). The origin of these reflection features can be due to the reprocessing of the primary X-ray continuum in the accretion disk and other neutral and ionized mediums around the central engine. The low-energy absorption is typically significant in Seyfert 2 AGNs, often due to a dusty torus, as described in the unification model (R. Antonucci 1993). In contrast, the Seyfert 1s are usually not absorbed or only marginally absorbed with absorption column density, $N_{\rm H} \leqslant 10^{22}$ cm$^{-2}$. The absence of strong absorption allows us to see an additional featureless spectral component, known as soft X-ray excess (below 2 keV) over the hard X-ray power-law continuum in many Seyfert 1s (J. P. Halpern 1984; K. A. Arnaud et al. 1985; T. J. Turner & K. A. Pounds 1989; G. A. Matzeu et al. 2020; Y. Xu et al. 2021; S. Chalise et al. 2022). However, the origin of this spectral component has been a topic of research over the last ∼40 yr.

---

[5] Corresponding author.

Table 1
Log of Observations of Mrk 50

| ID | Date (yyyy-mm-dd) | Obs. ID | Observatory | Instrument | Exposure (ks) |
|---|---|---|---|---|---|
| S1 | 2007-11-14 – 2007-11-16 | 00037089001 – 00037089003 | Swift | UVOT and XRT | 20.4 |
| X1 | 2009-07-09 | 0601781001 | XMM-Newton | OM, RGS, PN and MOS | 11.9 |
| X2 | 2010-12-09 | 0650590401 | XMM-Newton | RGS, PN and MOS | 22.9 |
| S2 | 2013-12-27 | 00080077001 | Swift | UVOT and XRT | 6.5 |
| S3 | 2022-06-27 | 00080077002 | Swift | UVOT, XRT and BAT | 5.7 |
| NU | 2022-06-27 | 60061227002 | NuSTAR | FPMA and FPMB | 17.4 |

It was initially believed that the soft excess originated from the inner part of the accretion disk as a hard-tail of the UV blackbody emission (K. P. Singh et al. 1985; K. A. Pounds et al. 1986; K. M. Leighly 1999a; P. Magdziarz et al. 1998; K. M. Leighly 1999b). However, this idea has been ruled out due to the high disk temperature (∼0.2 keV) of a typical AGN accretion disk (N. I. Shakura & R. A. Sunyaev 1973). In addition to that, different masses of the central black hole and different accretion rates for different AGNs indicate a wide range of disk temperature, which is inconsistent with the narrow range of disk temperature required to explain the soft excess observed in many sources (M. Gierliński & C. Done 2004; D. Porquet et al. 2004; E. Piconcelli et al. 2005; S. Bianchi et al. 2009; G. Miniutti et al. 2009). The present understanding of the origin of soft excess favors either the warm Comptonizing corona model (B. Czerny & M. Elvis 1987; M. Middleton et al. 2009; C. Done et al. 2012; A. Kubota & C. Done 2018; P. O. Petrucci et al. 2018; P. O. Petrucci et al. 2020) or the blurred ionized reflection model (A. C. Fabian et al. 2002; R. R. Ross & A. C. Fabian 2005; J. Crummy et al. 2006; J. Garcìa & T. R. Kallman 2010; D. J. Walton et al. 2013). In the case of warm Comptonization model, the optical/UV photons from the accretion disk are Compton upscattered in a warm ($kT_e \sim$ 0.1–1 keV) and optically thick ($\tau \sim$ 10–40) corona covering the inner part of the disk, producing excess emission below ∼2 keV (B. Czerny & M. Elvis 1987; C. Done et al. 2012; P. O. Petrucci et al. 2018; S. Tripathi et al. 2019; R. Middei et al. 2020; F. Ursini et al. 2020). In the other model, the relativistically blurred and ionized reflection model predicts that the soft excess component is generated by the reprocessed hard X-rays from the hot corona in the inner disk. This processed emission produces several fluorescent atomic lines, which are blended and distorted due to the strong gravity of the black hole at the inner edge of the accretion disk (J. Jiang et al. 2018; J. A. Garcìa et al. 2019).

Markarian 50 (Mrk 50) is a Seyfert 1 galaxy at a redshift of $z = 0.023$ (R. V. Vasudevan et al. 2013). Optical observations revealed that Mrk 50 exhibited significant variability in its nuclear region between 1985 and 1990, without any evidence of a disk structure in the system (M. G. Pastoriza et al. 1991). Mrk 50 is very bright in the X-ray range and known to be an unabsorbed Seyfert 1 galaxy with a soft excess and without any detectable iron line (R. V. Vasudevan et al. 2013). The first measurement of the black hole mass in Mrk 50 was determined (A. J. Barth et al. 2011) by using reverberation mapping technique and calculating the BH mass as $(3.2 \pm 0.5) \times 10^7 M_\odot$. Using the 2011 observation, A. Pancoast et al. (2012) reported the inclination angle ($i$) of Mrk 50 to be $9^{+7}_{-5}$ deg, indicating that the system is close to face-on. They also estimated the mass of the central black hole to be $3.71^{+0.44}_{-0.27} \times 10^7 M_\odot$. M. C. Bentz & S. Katz (2015) made a database of AGN black hole masses by compiling all published spectroscopic reverberation mapping studies of active galaxies and quoted the black hole mass in Mrk 50 as $3.55^{+0.45}_{-0.48} \times 10^7 M_\odot$.

This work presents our findings from a comprehensive study of Mrk 50 using X-ray observations from various X-ray satellites. This paper is organized as follows. Section 2 provides an overview of the observations and outlines the procedures used for data reduction. Detailed analyses of the temporal and spectral behaviors of the source are presented in Sections 3.1 and 3.2, respectively. Then, we discuss our key findings in Section 4, and finally, our conclusions are summarized in Section 5.

## 2. Observation and Data Reduction

Mrk 50 was observed with XMM-Newton (F. Jansen et al. 2001), Swift (N. Gehrels et al. 2004), and NuSTAR (F. A. Harrison et al. 2013) observatories at different epochs. The publicly available data are reduced and analyzed using the HEAsoft v6.30.1 package. The details of the observations used in the present work are given in Table 1.

### 2.1. Swift

Swift (N. Gehrels et al. 2004) is a multiwavelength satellite consisting of three onboard telescopes, operating in the optical/UV (Ultra-violet Optical Telescope, UVOT; P. W. A. Roming et al. 2005), X-ray (X-ray Telescope, XRT; D. N. Burrows et al. 2005), and hard X-ray (Burst Alert Telescope, BAT; S. D. Barthelmy et al. 2005) wave bands. Mrk 50 was observed with Swift at several epochs from 2007–2022, with the most recent observation on 2022 June 27 in simultaneous with NuSTAR (see Table 1). We extracted the light curves and spectra from the observed data by using the web tool "XRT product builder"[6] (P. A. Evans et al. 2009). Additional finer steps, such as pile-up correction on the given observation ID/IDs, are also applied while using the web tool. The tool processes and calibrates the data and produces final spectra and light curves of Mrk 50 in two modes, e.g., window timing (WT) and photon counting (PC) modes. In 2007, Swift observed Mrk 50 in three consecutive days from 2007 November 14–16 with XRT for exposure times of ∼7 ks, ∼4.6 ks, and ∼8.7 ks, respectively. Due to the short exposures, the signal-to-noise ratios during these observations were poor. The hardness ratios (ratio between count rates in the 1.5–10 keV and the 0.3–1.5 keV band) were also found to be unchanged during these observations. We, therefore, considered these three observations together and produced a combined observation (S1) for an exposure time of 20.4 ks.

---

[6] http://swift.ac.uk/user_objects/

The Swift/UVOT provides data in three optical filters (*V*, *B*, and *U*) and three UV filters ( UVW1, UVM2, and UVW2). We started our UVOT data reduction from the level II image files, performing photometry using the tool `UVOT-SOURCE`. To obtain the source counts, we assumed a circular region of 5″ radius centered at the source position, whereas a circular region with 20″ radius, away from the source position, was considered for background counts. We used the `UVOT2PHA` tool to create `XSPEC`-readable source and background spectra and used response files provided by the Swift team. In this work, we used only data from the UV filters (UVW1, UVM2, and UVW2) and excluded the optical (*V*, *B*, and *U*) data due to the contribution from the host galaxy and starburst in this band.

For Swift/BAT, we downloaded the hard X-ray spectrum of Mrk 50 from the 105 Month Swift/BAT All-sky Hard X-Ray Survey.[7] In addition, a response matrix appropriate for the BAT spectra was used in our work.

### 2.2. XMM-Newton

XMM-Newton (F. Jansen et al. 2001) observed Mrk 50 at two epochs in 2009 July and 2010 December. The details of the observations are mentioned in Table 1. Data from the European Photon Imaging Camera (EPIC), Reflection Grating Spectrometer (RGS), and Optical Monitor (OM) instruments are used in the present work. The EPIC includes three X-ray CCD cameras covering the 0.3–10 keV energy bandpass: two Metal Oxide Semiconductors (MOS1 and MOS2; M. J. L. Turner et al. 2001) and a p-n CCD (PN; L. Strüder et al. 2001). The two RGS detectors (RGS1 and RGS2) provide high-resolution spectroscopy over the 0.35–2 keV energy range (J. W. den Herder et al. 2001). The OM (K. O. Mason et al. 2001) provides photometric data from optical to UV bands (*U*, *B*, *V*, UVW2, UVM2, and UVW1).

We analyzed the Observation Data File using XMM-Newton Science Analysis System (`SASv18.0.0`) and updated calibration files as of 2019 October 23. We processed all of the EPIC data using the tasks `EMPROC` and `EPPROC` to obtain the calibrated and integrated event lists for the MOS and PN detectors, respectively. We filtered the data using the standard filtering criterion. Data during high background flaring events were removed before extracting the spectral products by creating and choosing appropriate good time intervals (GTI) using the task `TABGTIGEN`. In this work, we considered only the unflagged events with PATTERN ⩽ 4 and PATTERN ⩽ 12 for the PN and MOS detectors, respectively. The data were checked for photon pile-up using the task `EPATPLOT` and corrected accordingly. The source photons were extracted by considering an annular region with outer and inner radii of 30″ and 5″, respectively, centered at the source coordinates. We used a circular region of 60″ radius, away from the source position, for the background products. The response files for PN and MOS detectors were generated using SAS tasks `ARFGEN` and `RMFGEN` for *arf* and *rmf* files, respectively.

We reduced the RGS data using the standard SAS task `RGSPROC` and the updated calibration files to produce the source and background spectra. The cleaned event lists were generated by applying the standard filtering criteria. The response matrices were generated using the `RGSRMFGEN` task. The individual RGS1 and RGS2 spectra were combined into a single merged spectrum using the task `RGSCOMBINE` to achieve better signal-to-noise for the purpose of spectral fitting.

The Optical Monitor (OM; K. O. Mason et al. 2001) was also used simultaneously to observe Mrk 50 in 2009 in imaging mode with UVW1 and UVW2 filters. For the 2010 observation (0650590401), however, the optical/UV data from the OM were not available. We processed the OM data using the task `OMCHAIN`. The `om2pha` command was used to generate `XSPEC`-readable spectral files for all of the available filters. For the OM data, we used the canned response files available in the ESA XMM-Newton website.[8] We obtained background-corrected count rates from the source list for each OM filter and converted the count rates into respective flux densities.[9]

### 2.3. NuSTAR

NuSTAR is a hard X-ray focusing telescope consisting of two identical focal plane modules, FPMA and FPMB, and operates in the 3–79 keV energy range (F. A. Harrison et al. 2013). Mrk 50 was observed with NuSTAR simultaneously with Swift in 2022 June. The observation details are presented in Table 1. We use the standard NuSTAR Data Analysis Software (`NuSTARDASv2.1.2`[10]) package to extract data. The standard `NUPIPELINE` task with the latest calibration files CALDB[11] is used to generate the cleaned event files. The `NUPRODUCTS` task is utilized to extract the source spectra and light curves. We consider circular regions of radii 60″ and 120″ for the source and background products, respectively. The circular region for the source is selected with the center at the source coordinates, and the region for the background is chosen far away from the source to avoid contamination.

## 3. Data Analysis and Result

### 3.1. Timing Analysis

We conducted the timing analysis of Mrk 50 using light curves from the Swift/XRT, XMM-Newton, and NuSTAR observations (see Table 1). The time resolution of the light curves used in our analysis is 200 s. The light curves in the 0.3–10 keV range, generated from the XMM-Newton and Swift/XRT observations, and in the 3–60 keV range from NuSTAR are shown in Figure A1. Further, we extract light curves in the soft (0.3–3 keV range) and hard (3–10 keV range) X-ray bands for the variability and correlation studies. The light curve obtained from the NuSTAR observation is used to explore the variability of the source in the high-energy domain (3–60 keV). The entire energy range (3–60 keV) is further subdivided into two energy bands: band 1 (3–10 keV) and band 2 (10–60 keV), for variability studies.

#### 3.1.1. Fractional Variability

To check the temporal variability of Mrk 50 across different energy bands, we calculate the fractional variability $F_{var}$ (R. A. Edelson et al. 1996; K. Nandra et al. 1997; P. M. Rodríguez-Pascual et al. 1997; S. Vaughan et al. 2003; R. Edelson & M. Malkan 2012). The fractional variability for a light curve of $x_i$ counts s$^{-1}$ with the measurement error $\sigma_i$ for *N* number of

---

[7] https://swift.gsfc.nasa.gov/results/bs105mon/
[8] https://sasdev-xmm.esac.esa.int/pub/ccf/constituents/extras/responses/OM/
[9] https://www.cosmos.esa.int/web/xmm-newton/sas-watchout-uvflux
[10] https://heasarc.gsfc.nasa.gov/docs/nustar/analysis/
[11] http://heasarc.gsfc.nasa.gov/FTP/caldb/data/nustar/fpm/

Table 2
Variability Statistics in Different Energy Ranges for Various Observations, Calculated Using Light Curves with 200 s Time Bin

| ID | Energy (keV) | N | $x_{\max}$ (count s$^{-1}$) | $x_{\min}$ (count s$^{-1}$) | $\mu$ (count s$^{-1}$) | $R = \frac{x_{\max}}{x_{\min}}$ | $\sigma^2_{\mathrm{NXS}}$ ($10^{-2}$) | $F_{\mathrm{var}}$ (%) |
|---|---|---|---|---|---|---|---|---|
| S1 | 0.3–3 | 122 | 0.87 | 0.19 | 0.50 | 4.57 | 5.17 ± 1.07 | 22.74 ± 2.78 |
|  | 3–10 | 118 | 0.31 | 0.02 | 0.09 | 14.44 | 3.77 ± 4.54 | 19.42 ± 11.77 |
|  | 0.3–10 | 122 | 1.04 | 0.23 | 0.58 | 4.47 | 4.31 ± 1.81 | 20.77 ± 2.68 |
| X1 | 0.3–3 | 50 | 9.87 | 8.30 | 9.01 | 1.20 | 0.05 ± 0.02 | 2.26 ± 0.56 |
|  | 3–10 | 50 | 0.72 | 0.39 | 0.55 | 1.87 | 0.30 ± 0.31 | 5.49 ± 2.95 |
|  | 0.3–10 | 51 | 10.38 | 8.79 | 9.55 | 1.18 | 0.04 ± 0.02 | 2.00 ± 0.60 |
| X2 | 0.3–3 | 105 | 5.78 | 4.25 | 5.02 | 1.36 | 0.27 ± 0.04 | 4.79 ± 0.56 |
|  | 3–10 | 106 | 0.64 | 0.25 | 0.42 | 2.52 | 0.53 ± 0.40 | 7.27 ± 2.78 |
|  | 0.3–10 | 104 | 6.21 | 4.63 | 5.44 | 1.34 | 0.23 ± 0.04 | 4.87 ± 0.05 |

**Note.** In most of the cases (S2, S3, and NU observations), the average error of the observational data surpasses the 1σ limit, resulting in negative excess variance. As a result, these cases contain imaginary values for $F_{\mathrm{var}}$, and thus, they are excluded from the Table.

data points, mean count rate $\mu$, and standard deviation $\sigma$, is given by the relation:

$$F_{\mathrm{var}} = \sqrt{\frac{\sigma^2_{\mathrm{XS}}}{\mu^2}} \quad (1)$$

where, $\sigma^2_{\mathrm{XS}}$ is the excess variance (K. Nandra et al. 1997; R. Edelson et al. 2002), used to estimate the intrinsic source variance and given by,

$$\sigma^2_{\mathrm{XS}} = \sigma^2 - \frac{1}{N}\sum_{i=1}^{N} \sigma_i^2. \quad (2)$$

The normalized excess variance is defined as $\sigma^2_{\mathrm{NXS}} = \sigma^2_{\mathrm{XS}}/\mu^2$. The uncertainties in $\sigma^2_{\mathrm{NXS}}$ and $F_{\mathrm{var}}$ are estimated as described in S. Vaughan et al. (2003) and R. Edelson & M. Malkan (2012). The peak-to-peak amplitude is defined as $R = x_{\max}/x_{\min}$ (where, $x_{\max}$ and $x_{\min}$ are the maximum and minimum count rates, respectively) to investigate the variability in the X-ray light curves.

During the 2007 observation (S1), we found that the average count rate $\mu$ and the peak-to-peak amplitude $R$ remain constant at ∼0.55 count s$^{-1}$ and ∼4.50 for both the soft X-ray and the entire energy bands, respectively. In these energy bands, the normalized excess variance ($\sigma^2_{\mathrm{NXS}}$) and corresponding fractional variability ($F_{\mathrm{var}}$) are found to be constant at ∼0.05% and ∼22%, respectively. For the hard X-ray band, the average count rate, $\mu$, decreases to 0.08 count s$^{-1}$, and the corresponding $R$ value increases to 14.44. In this energy band, the fractional variability ($F_{\mathrm{var}}$) is estimated to be 19.42% ± 11.77%. So, in this epoch (S1), the source was variable (∼20%) in the soft X-ray, hard X-ray, and the entire energy band. The average source count rate was maximum during the 2009 and 2010 XMM-Newton observations (X1 and X2) compared to the epochs of other observations used in the present work. While estimating the fractional variability ($F_{\mathrm{var}}$), we find that in all of the energy bands, $F_{\mathrm{var}}$ is < 10%. This indicates that the source was nonvariable during these observations. For S2, S3, and NU observations, the variability parameters like $\sigma^2_{\mathrm{NXS}}$ and $F_{\mathrm{var}}$ are also calculated. However, due to the low count rate and high error associated with each data point, we encounter negative values for normalized excess variance, resulting in imaginary fractional variability. The details of the results of the variability analysis are presented in Table 2.

### 3.1.2. Cross-correlation

We carried out the cross-correlation function (CCF) analysis to search for a correlation and time lag between X-ray light curves in different energy bands. We applied the $\zeta$-transformed discrete correlation function ($\zeta - \mathrm{DCF}$[12]) method (T. Alexander 1997, 2013) to estimate the CCF between the light curves. This method is suitable for both evenly and unevenly sampled data but is particularly appropriate when the observed data are unevenly sampled and sparse. However, the ZDCF binning algorithm uses a logic similar to that of the discrete correlation function (DCF) developed by R. A. Edelson & J. H. Krolik (1988). However, the ZDCF method adopts equal population binning and Fisher's z-transform to correct several biases of the DCF method. We used the available FORTRAN95[13] code to carry out the cross-correlation. While performing the ZDCF analysis, we used 102,000 Monte Carlo runs for the error estimation of the coefficients. Also, we did not consider the points for which the lag was zero. The X-ray light curves in different energy bands from different observations are shown in the upper panels of Figure 1. Furthermore, we plot the correlation function between the light curves of different energy bands in the middle panels of Figure 1. The count–count plots are also presented in the bottom panels of the same Figure.

To probe the origin of the soft excess in Mrk 50, we investigate the time delay between the soft X-ray (0.3–3 keV) and hard X-ray (3–10 keV) bands using the cross-correlation method. We begin our analysis using data from observation S1 and find that the soft and hard X-ray bands are uncorrelated. We find similar results in the case of observations X1 and X2. We do not notice any significant peak in the $\zeta$ − discrete cross-correlation function in any of the observations (see Figure 1). The detection of no correlation between soft and hard X-ray bands suggests that the photons in both energy bands could have originated through different physical mechanisms. We then investigate the correlation between the light curves in the high-energy regime. Light curves from the NU observations in band1 and band2 are utilized to explore the correlation in the high-energy regime. In this case, we find that these two bands are uncorrelated. The details of our findings from the cross-correlation study are explained in Section 4.4.2.

---
[12] www.weizmann.ac.il/particle/tal/research-activities/software
[13] https://www.weizmann.ac.il/particle/tal/research-activities/software

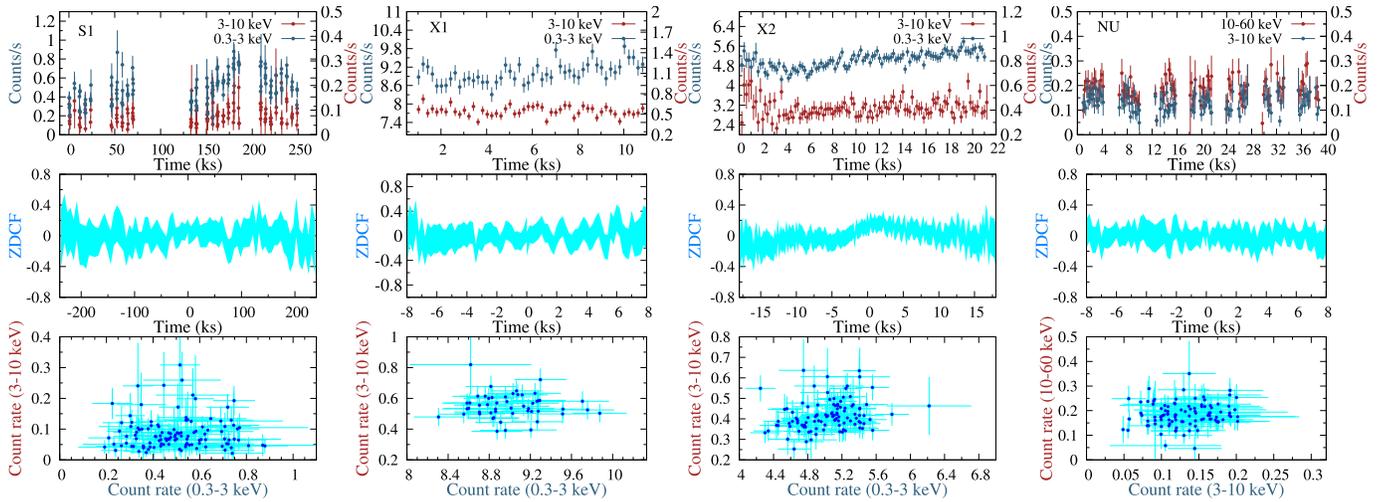

**Figure 1.** Light curves in different energy ranges (top panels), correlation between the corresponding light curves (middle panels), and the count–count plot (bottom panels) from three different X-ray instruments (Swift/XRT, XMM-Newton, and NuSTAR) are shown.

### 3.2. Spectral Analysis

The spectral analysis is carried out using Swift (XRT and UVOT) and XMM-Newton (EPIC-pn, MOS & OM) observations in 0.001–10 keV range and simultaneous Swift (UVOT, XRT, and BAT) and NuSTAR observations in 0.001–120 keV range (see Table 1). The NuSTAR data beyond 60 keV are not considered in the present analysis as it is dominated by background. We use the XSPEC v12.12.1 (K. A. Arnaud 1996) software package for spectral fitting. We binned the spectral data at a minimum of 25 counts/bin for both XMM-Newton and NuSTAR observations and 20 counts in each bin for the Swift/XRT observations. The GRPPHA task is used for binning the spectral data. The model likelihoods are determined using the $\chi^2$ statistic. However, it is to be noted that the Swift/BAT data are Gaussian,[14] and we continue to present the $\chi^2$-statistic here for simplicity. We also ignored the bad channels in our analysis.

The best-fit parameters are reported in the rest frame of the source. The uncertainties corresponding to the model parameters are quoted at the 90% confidence level using the Monte Carlo Markov Chain (MCMC) method embedded in XSPEC. The MCMC technique simultaneously determines the errors in the model parameters and renders better parameter space sampling than other methods. We used the Goodman & Weare sampler (J. Goodman & J. Weare 2010) to deal with degeneration in the model parameters. For sampling, we choose the number of walkers to be more than twice the number of free model parameters. In all cases, we consider the chain length of 200,000 for the chains to converge in the same parameter values. The first 10,000 steps are discarded for the burn-in period to remove the bias introduced by the choice of the starting location. To ensure that the walkers have enough sampling in the parameter space, we checked whether the steps were rejected <75% of the time or not.

The unabsorbed X-ray luminosity from each spectrum is estimated using clumin task on the powerlaw model. While estimating the luminosity, we use the redshift, $z = 0.023$. We also calculate the X-ray flux by using cflux command in XSPEC. The observed UV flux is corrected for reddening and Galactic extinction using the reddening coefficient $E(B - V) = 0.0147$ obtained from the Infrared Science Archive[15] and $R_V = A_V/E(B - V) = 3.1$ following E. F. Schlafly & D. P. Finkbeiner (2011). The Galactic extinction ($A_\lambda$) value used is 0.140. Throughout this work, we use the Cosmological parameters as follows: $H_0 = 70$ km s$^{-1}$ Mpc$^{-1}$, $\Lambda_0 = 0.73$, and $\sigma_M = 0.27$ (C. L. Bennett et al. 2003).

### 3.3. Characterizing the Spectrum

As Mrk 50 has not been explored in the high-energy domain to date, our first motivation is to characterize the spectrum of this source. For this purpose, we begin our spectral fitting with a set of phenomenological models to characterize the source spectra at different epochs and determine the spectral features quantitatively. Further, these phenomenological models are replaced with more sophisticated physical models to better understand the physical properties of the source.

Initially, we consider the 3–10 keV X-ray continuum spectra of the source for the spectral fitting. According to current understanding, the X-ray continuum photons are produced through the inverse Compton scattering of thermal photons from the accretion disk in a hot electron cloud (N. I. Shakura & R. A. Sunyaev 1973), though the geometry and location of this hot electron cloud are poorly understood. This nonthermal process leads to a power-law-type spectrum. Therefore, we applied a simple Powerlaw model to fit the spectrum of each observation. Along with this, we use the Galactic line-of-sight hydrogen column density ($N_H$, Gal) as the multiplicative model TBabs (J. Wilms et al. 2000) in XSPEC. The value of $N_H$, Gal[16] used in our work is $1.71 \times 10^{20}$ cm$^{-2}$. We note that we did not observe any prominent (positive) residual in the 6–7 keV energy range during the continuum fitting of any of the five sets of spectra (see left panel of Figure 2). This indicates the absence of the Fe-line in the X-ray continuum of Mrk 50. So, for the continuum, the baseline model is Const × TBabs × Powerlaw for all observations. The Constant component is used as a cross-normalization factor while using data from different instruments in simultaneous spectral fitting (see Table A1).

After successfully parameterizing the primary continuum in the 3–10 keV range, we extend the spectrum into the low-energy domain (below 3 keV), where we find distinct results across different epochs of observations (see right panel of Figure 2). In the case of S1, we notice deviation in the low-energy data points from the primary continuum, which is attributed to the presence of soft excess at $E < 2$ keV. To address the presence of soft excess in this observation, we consider another power-law component below 3 keV (R. Walter & H. H. Fink 1993; P. Nandi et al. 2021). To ensure the significance of this additional Powerlaw component, we conducted an F-test[17] for the additive component, and find the $F_{\text{value}}$ to be 9.86 with a probability of chance improvement $1.40 \times 10^{-5}$. Using the F-test, we verified that an additional component is required. We also find evidence of soft excess in X1 and X2 observation.

---

[14] https://swift.gsfc.nasa.gov/analysis/threads/batspectrumthread.html

[15] http://irsa.ipac.caltech.edu/applications/DUST/
[16] https://heasarc.gsfc.nasa.gov/cgi-bin/Tools/w3nh/w3nh.pl

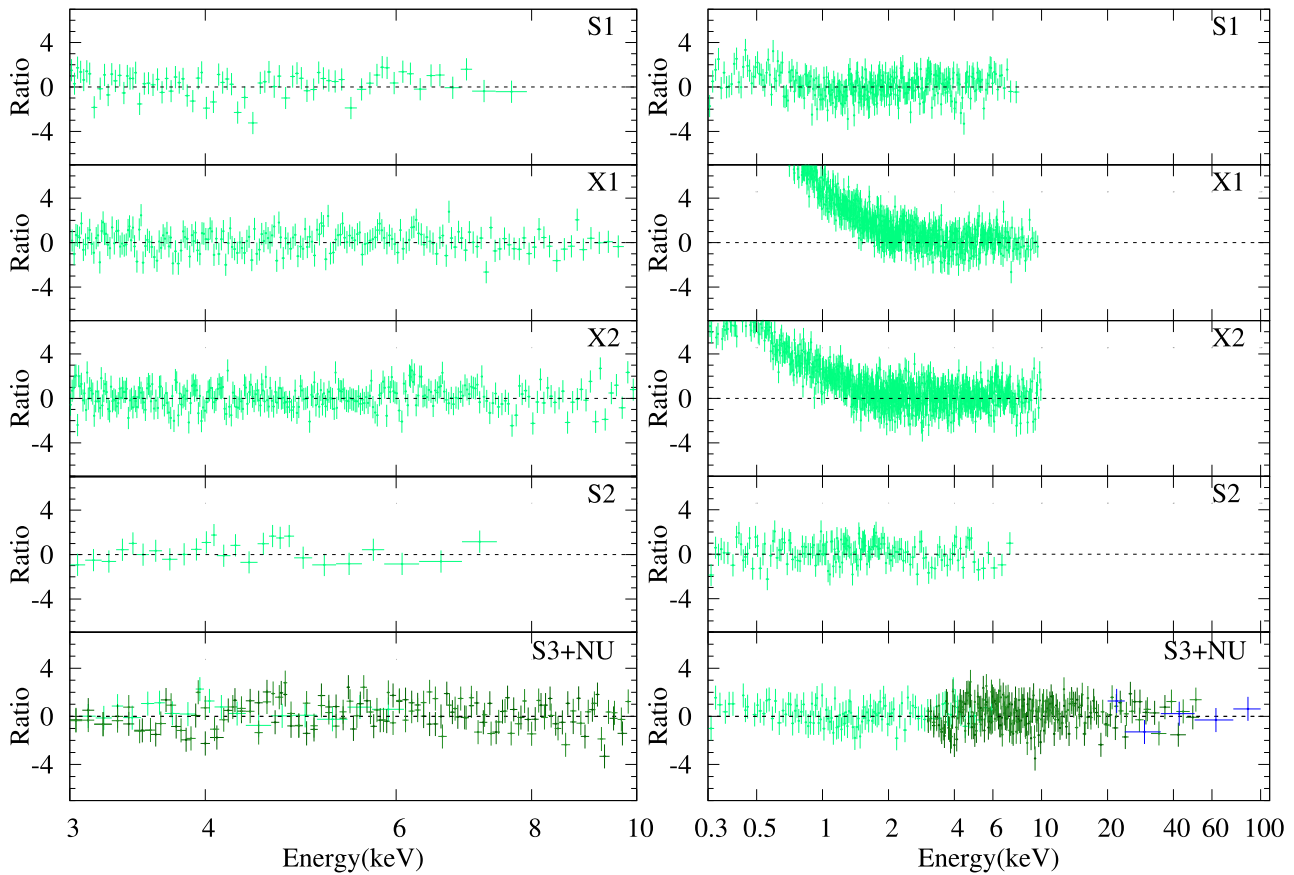

**Figure 2.** The ratio plots of five spectra against Galactic absorbed power-law models. None of the spectra show the presence of any iron emission line in the iron band (6–7 keV). A strong soft excess is observed in the spectra of 2007 (S1), 2009 (X1), and 2010 (X2) observations. However, no soft excess was detected in the spectra of 2013 (S2) and 2022 (S3+NU) observations.

However, for observations S2 and S3+NU, it is surprising that the low-energy spectra (below 3 keV) do not show the presence of any excess emission over the continuum. The primary continuum fits the extended low-energy range of these observations. The exposure times for S2 and S3 observations are shorter compared to the other observations, which may explain the nondetection of the soft excess component below 3 keV. However, for further clarification, we performed an F-test on these spectra. The F-test confirms that adding the extra power-law component is insignificant and did not improve the fit.

To investigate the presence of any intrinsic absorption along the line of sight of Mrk 50, we initially used the neutral absorption model component, zTbabs, with our baseline model and found that the estimated value of $N_{\text{H}}$ is $< 0.001 \times 10^{22}$ cm$^{-2}$ for $z = 0.024$. To investigate further, we replace zTbabs with a partially covering absorption model Pcfabs to check for the presence of any partial absorbers present in the line of sight. However, this model is also found to be insensitive during spectral fitting and consistently yields the lowest value of the absorption parameter. We considered UV data for these observations in our spectral fitting to ensure our findings. However, the inclusion of UV data in the fitting did not detect the presence of any extragalactic absorption component in Mrk 50. As a result, we drop this zTbabs component from our composite model. The corresponding model in XSPEC reads as Const × TBabs × (powerlaw + powerlaw) and is used to fit all of the spectra in the 0.3–10 keV range. Later, we extend the low-energy part into the UV domain whenever the UV observations are available.

In the high-energy domain (>10 keV), we combine spectra from the Swift/XRT, NuSTAR, and Swift/BAT to obtain a broadband X-ray spectrum up to 100 keV. While extrapolating the primary continuum model to the high-energy range, we did not find any deviation in the data points from the primary model (see Figure 2). This suggests that the reflection component in the X-ray spectra above 10 keV is either absent or insignificant during this observation.

---
[17] https://heasarc.gsfc.nasa.gov/xanadu/XSpec/manual/node82.html

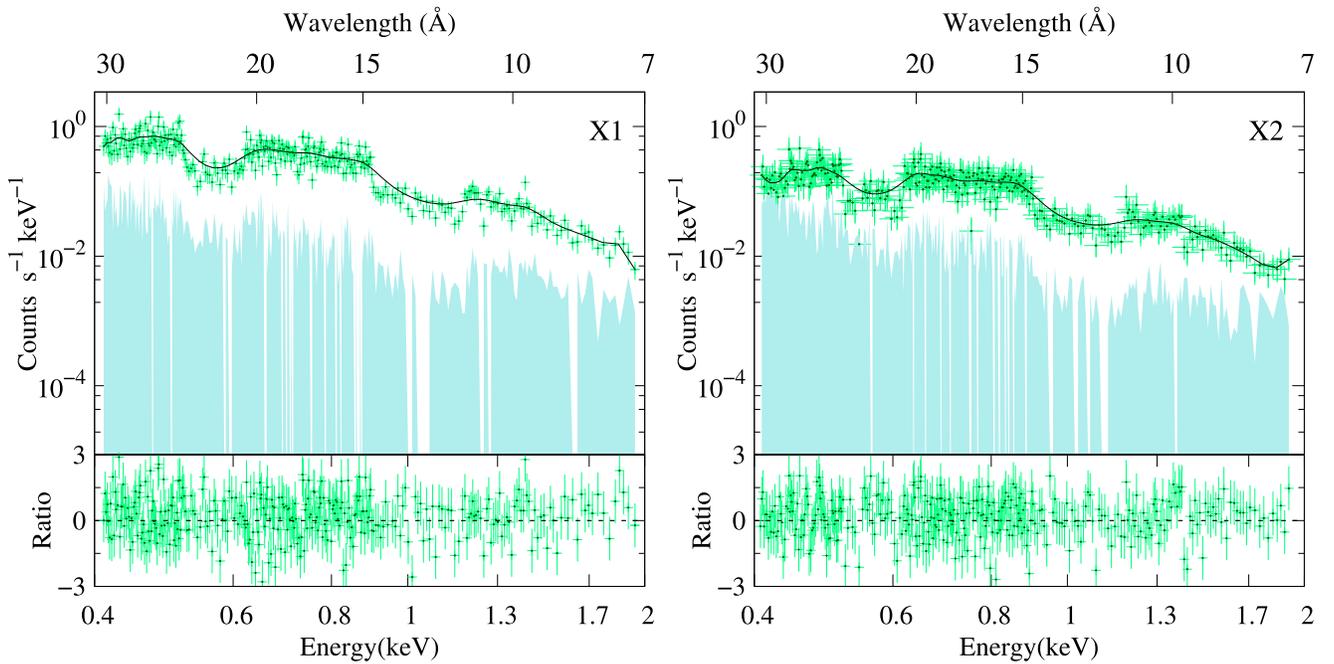

**Figure 3.** The merged RGS (RGS1+2) spectrum from 2009 and 2010 XMM-Newton observations of Mrk 50. The best-fit simple power-law model modified by the Galactic absorption is overlaid in a solid black line, and the background level is shown in cyan. The bottom panel illustrates the corresponding residuals against the absorbed power-law model.

After characterizing each spectrum of all of the observations with simple models, such as a power law, we use the phenomenological model `Diskbb` to address the soft X-ray excess (see Section 3.3.3) in this source. To better understand the physical nature of the soft X-ray excess emission in Mrk 50, we use `Optxagnf` and `Relxillcp` as physical models. The results are discussed in the Sections that follow.

### 3.3.1. RGS Spectral Analysis

Before proceeding to a detailed spectral analysis of the source, we first present a brief analysis of the RGS data to confirm the "bare" nature of Mrk 50. We analyzed the merged RGS spectrum (RGS1+2) of Mrk 50 to check for possible prominent soft X-ray absorption and/or emission in the energy range of 0.4–2 keV. We adopt the spectral binning of 20 counts/bin and use the $\chi^2$ statistics. The merged RGS spectrum is initially modeled by simple `powerlaw` modified by the Galactic absorption. The absorbed power-law model provided a good fit with the $\chi^2$-statistic of 407.52 for 395 degrees of freedom and $\Gamma$ of 2.38 ± 0.05 for the X1 observation. For X2 observation, we found a $\chi^2$-statistic of 299.43 for 320 degrees of freedom and $\Gamma$ of 2.10 ± 0.06. Figure 3 shows the absorbed power-law fitted merged RGS spectra of Mrk 50 for both the XMM-Newton observations (X1 and X2), with the residual shown in the lower panel. To investigate the presence of any ionized absorption, we used the multiplicative model `Zxipcf` for the power-law model. This addition resulted in an insignificant improvement in the fit statistics. The upper limit of the column density was constrained to $N_H < 6.33 \times 10^{-20}$ cm$^{-2}$ for X1 observation. However, this model was also found to be insensitive during spectral fitting of X2 observation. Such a lower value of the $N_H$ was previously seen in other systems (S. Laha et al. 2014; G. A. Matzeu et al. 2020; A. Madathil-Pottayil et al. 2024; P. Nandi et al. 2024; D. Porquet et al. 2024). This indicates that no significant warm absorbing gas exists along our line of sight toward this AGN, thereby supporting its classification as a bare Seyfert 1 galaxy.

### 3.3.2. Powerlaw

We started our spectral analysis with an absorbed power-law model as described in Section 3.3. From the spectral fitting, we found that the power-law indices of the primary continuum ($\Gamma_{PC}$) varied from $1.54^{+0.42}_{-0.40}$ to $1.88^{+0.27}_{-0.28}$, and the corresponding luminosities (log$L_{PC}$) varied from $43.14^{+0.02}_{-0.06}$ to $43.58^{+0.02}_{-0.04}$. After the primary continuum fitting, we fitted the lower energy range (below 3 keV) of the observed spectrum using another power-law component (see Section 3.3). We considered available UV data in the spectral fitting, and used the multiplicative model `Redden` (J. A. Cardelli et al. 1989) with a fixed value of $E(B - V) = 0.0147$ (E. F. Schlafly & D. P. Finkbeiner 2011) to account for the interstellar extinction. The model for spectral fitting of data in the 0.001–10 keV range is represented in XSPEC as `Const × TBabs × Redden × (Powerlaw + Powerlaw)`.

For the 2007 observation (S1), we found the power-law indices of soft excess ($\Gamma_{SE}$) and the primary continuum ($\Gamma_{PC}$) are in the same order. This indicates a minimum presence of soft excess in this observation period. We also calculated the corresponding luminosities and found that these values are nearly the same. The primary continuum luminosity ($L_{PC}$) for S1 is calculated as $3.8^{+0.2}_{-0.4} \times 10^{43}$ erg s$^{-1}$, whereas the soft excess luminosity ($L_{SE}$) is $4.1^{+0.1}_{-0.5} \times 10^{43}$ erg s$^{-1}$. From X1 (2009) and X2 (2010) XMM-Newton observations, we found a difference in $\Gamma_{PC}$ and $\Gamma_{SE}$ along with the luminosities, indicating a strong presence of soft excess in these two observations. The deviation from the primary power-law continuum below 3 keV is significant in X1 (see Figure 2), which is characterized by $\Gamma_{SE} = 2.63 \pm 0.04$. In X2, the amount of soft excess is reduced, but a significant presence of this component is still observed with $\Gamma_{SE} = 2.58^{+0.07}_{-0.05}$.

**Table 3**
Best-fit Parameters of the Baseline Phenomenological Model Constant × Tbabs × Redden × (powerlaw + powerlaw) for the Mrk 50 Observations.

| ID | $\Gamma_{PC}$ | Norm$^a_{PC}$ ($10^{-3}$) | log$L_{PC}$ log(erg s$^{-1}$) | $\Gamma_{SE}$ | Norm$^a_{SE}$ ($10^{-3}$) | log$L_{SE}$ log(erg s$^{-1}$) | $\chi^2$/dof |
|---|---|---|---|---|---|---|---|
| S1 | $1.88^{+0.27}_{-0.28}$ | $3.54^{+0.17}_{-0.11}$ | $43.58^{+0.02}_{-0.04}$ | $1.91^{+0.03}_{-0.04}$ | $23.28^{+4.70}_{-3.94}$ | $43.61^{+0.01}_{-0.06}$ | 260.93/203 |
| X1 | $1.72 \pm 0.05$ | $3.11^{+0.28}_{-0.25}$ | $43.35^{+0.02}_{-0.01}$ | $2.63 \pm 0.04$ | $2.36^{+0.11}_{-0.10}$ | $43.64^{+0.02}_{-0.01}$ | 1864.48/1597 |
| X2 | $1.70 \pm 0.05$ | $2.12^{+0.17}_{-0.15}$ | $43.38^{+0.03}_{-0.02}$ | $2.58^{+0.07}_{-0.05}$ | $0.80^{+0.07}_{-0.06}$ | $43.15^{+0.02}_{-0.01}$ | 1894.17/1725 |
| S2 | $1.54^{+0.42}_{-0.40}$ | $1.94^{+1.62}_{-0.89}$ | $43.46^{+0.21}_{-0.36}$ | $1.80^{+0.24}_{-0.21}$ | $3.96^{+0.04}_{-1.10}$ | $43.18^{+0.15}_{-0.16}$ | 73.95/84 |
| S3+NU | $1.77^{+0.05}_{-0.07}$ | $2.43^{+0.04}_{-0.03}$ | $43.14^{+0.02}_{-0.06}$ | $1.77^p$ | $16.34^{+1.16}_{-1.10}$ | < 42.91 | 238.61/196 |

**Notes.** Spectral fitting of all observations includes simultaneous optical-UV data except X2. The soft excess (SE) and power-law continuum (PC) luminosity are calculated in the energy range 0.001–10 keV.
[a] In units of photons keV$^{-1}$ cm$^{-2}$ s$^{-1}$.

**Table 4**
Best-fit Parameters of the Baseline Phenomenological Model Diskbb for the Observations of Mrk 50

| ID | $T_{in}$ (keV) | Norm$_{BB}$ ($10^2$) | $\Gamma$ | Norm$^a_{PL}$ ($10^{-3}$) | $\chi^2$/dof |
|---|---|---|---|---|---|
| S1 | $0.16^{+0.03}_{-0.01}$ | $86.96^{+44.87}_{-34.20}$ | $1.68^{+0.07}_{-0.06}$ | $64.43^{+1.80}_{-2.63}$ | 213.50/202 |
| X1 | $0.14 \pm 0.01$ | $8.15^{+1.77}_{-1.36}$ | $1.91 \pm 0.02$ | $4.09^{+0.07}_{-0.06}$ | 1574.47/1596 |
| X2 | $0.18 \pm 0.01$ | $0.97^{+0.29}_{-0.18}$ | $1.76 \pm 0.02$ | $2.31^{+0.06}_{-0.05}$ | 1836.74/1724 |
| S2 | $0.10^b$ | >11.14 | $1.78^{+0.05}_{-0.07}$ | $2.80^{+0.11}_{-0.12}$ | 77.19/84 |
| S3+NU | $0.10^b$ | >6.79 | $1.74^{+0.02}_{-0.03}$ | $23.22^{+5.10}_{-4.22}$ | 322.16/282 |

**Notes.** Spectral fitting of all observations includes simultaneous optical-UV data except X2.
[a] In units of photons keV$^{-1}$ cm$^{-2}$ s$^{-1}$.
[b] Indicates a frozen parameter.

After 3 yr of XMM-Newton observations, Mrk 50 was observed with Swift in 2013 (S2) in the UV and X-ray bands. In this observation (S2), we could not detect the soft excess emission below 3 keV. As a result, the values of power-law indices are found to be in the same order (within uncertainties). $\Gamma_{SE}$ and $\Gamma_{PC}$ are $1.80^{+0.24}_{-0.21}$ and $1.54^{+0.42}_{-0.40}$, respectively, for this observation.

In 2022 (S3+NU), a broadband simultaneous observation of the AGN was carried out with Swift (UVOT, XRT, and BAT) and NuSTAR. Although the spectrum is extended to 100 keV, initially, we considered data up to 10 keV from Swift (UVOT and XRT) and NuSTAR for the powerlaw+powerlaw model fitting. During the spectral fitting, the soft excess was barely detectable (see Figure 2), for which we could not constrain the power-law index and corresponding luminosity of the soft excess. We fixed $\Gamma_{SE}$ at 1.77, the same as $\Gamma_{PC}$, and corresponding luminosities are estimated to be $0.8 \times 10^{43}$ erg s$^{-1}$ and $1.38^{+0.06}_{-0.18} \times 10^{43}$ erg s$^{-1}$ for $L_{SE}$ and $L_{PC}$, respectively. It is to be noted that we are unable to calculate the errors in soft excess luminosity, as the $\Gamma_{SE}$ is fixed at 1.77. The values of the parameters obtained from spectral fitting are presented in Table 3.

### 3.3.3. Diskbb

After parameterizing the soft excess component by a simple power-law component, we started characterizing this excess using a multitemperature accretion disk blackbody model Diskbb (K. Mitsuda et al. 1984). The Diskbb is a relatively simple model with only two model parameters: the temperature at the inner edge of the disk $T_{in}$ and the model normalization. This model describes the data well in the optical/UV band. Therefore, we used simultaneous UV and X-ray data to investigate the soft excess in UV and X-ray spectra. For the spectral fitting, the model in XSPEC reads as Const × TBabs × Redden × (Diskbb + Powerlaw).

From the S1, we found that the inner disk temperature ($T_{in}$) is $0.16^{+0.03}_{-0.01}$ keV with a power-law index of $1.68^{+0.07}_{-0.06}$. A similar approach is followed for the spectral fitting of data from X1 and X2. As the soft excess is strong in these observations, we are able to constrain the inner disk temperature $T_{in}$. We found that the inner disk temperatures, ($T_{in}$), are $0.14 \pm 0.01$ and $0.18 \pm 0.01$ for X1 and X2, respectively. The power-law indices ($\Gamma$) are calculated as $1.91 \pm 0.02$ and $1.76 \pm 0.02$ for the corresponding observations.

In the case of S2 and S3+NU, as the presence of soft excess is below the limit of detection, we are unable to draw a limit on the inner disk temperature ($T_{in}$) of the Diskbb component. So, we fixed the value of $T_{in}$ at 0.10 keV. As the values of $T_{in}$s are not constrained to a certain limit, we are unable to limit the normalization of this model component. For the other model component (Powerlaw), we found the photon indices ($\Gamma$) are $1.80^{+0.05}_{-0.07}$ and $1.74^{+0.02}_{-0.03}$ for the S2 ad S3+NU, respectively. The values of the best-fit parameters and their fit statistics are quoted in Table 4.

### 3.4. The Physical Models

In this Section, our motivation is to understand the physical origin of each spectral component and its evolution over time. The origin of the soft excess component is yet to be understood. To explore this, we considered two possible physical scenarios, warm Comptonization (C. Done et al. 2012) and relativistic blurred reflection (J. A. García et al. 2018), on

Table 5
Best-fit Parameters for Observations of Mrk 50 with the Physical Model `Optxagnf`

| Models | Parameter | S1 | X1 | X2 | S2 | S3+NU |
|---|---|---|---|---|---|---|
| `Optxagnf` | $\log\lambda_{\rm Edd}$ | $-0.87 \pm 0.03$ | $-1.62 \pm 0.01$ | $-1.69^{+0.03}_{-0.02}$ | $-1.77^{+0.05}_{-0.03}$ | $-1.20^{+0.03}_{-0.05}$ |
| | $kT_e$ | $0.13^{+0.02}_{-0.01}$ | $0.13 \pm 0.01$ | $0.25^{+0.03}_{-0.02}$ | $0.10^a$ | $0.10^a$ |
| | $\tau$ | $29^{+4}_{-3}$ | $36^{+3}_{-2}$ | $18 \pm 3$ | $<5$ | $<3$ |
| | $R_{\rm cor}$ | $62^{+19}_{-9}$ | $98^{+1}_{-3}$ | $50^{+31}_{-20}$ | $<48$ | $<10$ |
| | $f_{\rm pl}$ | $0.88^{+0.02}_{-0.01}$ | $0.87^{+0.02}_{-0.01}$ | $0.82^{+0.04}_{-0.09}$ | $>0.95$ | $>0.93$ |
| | $\Gamma$ | $1.73 \pm 0.04$ | $1.92 \pm 0.01$ | $1.72 \pm 0.02$ | $1.79^{+0.04}_{-0.03}$ | $1.76^{+0.02}_{-0.04}$ |
| | $\chi^2$/dof | 202.18/198 | 1598.54/1594 | 1823.83/1722 | 68.90/80 | 323.59/279 |

**Note.** Spectral fitting of all observations includes simultaneous optical-UV data except X2.
[a] Indicates a frozen parameter.

the observed spectra to investigate the physical origin of this component.

### 3.4.1. Warm Comptonization

In this scenario, we used `Optxagnf` (C. Done et al. 2012) model as the warm Comptonization model to investigate the origin of soft excess in this source. It is an intrinsic thermal Comptonization model that describes the optical/UV emission of AGNs as multicolor blackbody from a color temperature-corrected disk. In this model, the disk emission emerges at radii $R_{\rm out} > r > R_{\rm cor}$, where $R_{\rm out}$ and $R_{\rm cor}$ are the outer edge of the disk and the corona, respectively. At $r < R_{\rm cor}$, the disk emission emerges as the Comptonized emission from a warm ($kT_e \sim 0.1$–1.0 keV) and optically thick ($\tau \sim 10$–40) plasma, expressed as the soft X-ray emission (P. Magdziarz et al. 1998; C. Done et al. 2012). The hot and optically thin corona is considered to be located around the disk and produces the high-energy power-law continuum. The observed Comptonized emission, therefore, consists of contributions from the cold and hot corona, with the fraction of hot-Comptonized emission determined by a parameter, $f_{\rm pl}$, obtained from the model fitting.

This model characterizes the total emission based on the mass accretion rate and the black hole mass. The soft X-ray excess emission is determined by parameters such as the temperature of the warm corona ($kT_e$), the temperature of the seed photon, and the optical depth of the warm corona ($\tau$) at $r = R_{\rm cor}$. The power-law continuum is approximated as the `nthcomp` model, with the seed photon temperature fixed at the disk temperature at $r = R_{\rm cor}$ and the electron temperature fixed at 100 keV. Four parameters that determine the model flux are the black hole mass ($M_{\rm BH}$), the Eddington ratio ($\lambda_{\rm Edd}$), the moving distance ($D$ in Mpc), and the dimensionless black hole spin ($a$). While using the `Optxagnf` model, we kept the black hole mass of Mrk 50 fixed at $3.55 \times 10^7 M_\odot$ (M. C. Bentz & S. Katz 2015) and the cosmological distance at 103 Mpc. As recommended, we fixed normalization to unity during our analysis. The warm Comptonization model uses the disk UV photons to produce the power-law continuum spectrum and the soft excess. Therefore, to constrain the `Optxagnf` parameters, we fitted the X-ray data along with simultaneously obtained UV data from the XMM-Newton and Swift observations, except for X2, where no UV data was available. The `Redden` model in XSPEC accounts for galactic extinction correction. During our analysis, the black hole spin parameter could not be well constrained when kept as a free parameter. Hence, we fixed the spin value at the maximal spin scenario ($a = 0.998$) in the fitting process.

From the spectral fitting of S1 data, we found that the accretion rate ($\log\lambda_{\rm Edd}$) is $-0.87 \pm 0.03$, with corresponding warm corona temperature ($kT_e$) and optical depth ($\tau$) of $0.13^{+0.03}_{-0.02}$ keV and $29^{+4}_{-3}$, respectively. In this observation, the soft excess emission is not prominent, and we found the coronal boundary ($R_{\rm cor}$) to be at $62^{+19}_{-9}$ $R_g$. The hard X-ray part of this spectrum is characterized by a power law with index ($\Gamma$) of $1.73 \pm 0.04$. The corresponding fraction of the energy below $R_{\rm cor}$ emitted in the hot corona is $f_{\rm pl} = 0.88^{+0.02}_{-0.01}$, while the fraction emitted in the warm corona is $(1 - f_{\rm pl}) = 0.12$. In the 2009 observation (X1), the accretion rate decreased to $-1.62 \pm 0.01$, leading to a warm corona temperature of $0.13 \pm 0.01$ keV and an optical depth of $36^{+3}_{-2}$. As the soft excess is prominent in this observation, we found that $R_{\rm cor}$ gets extended up to $98^{+1}_{-3}$ $R_g$. From the spectral fitting, we determined $\Gamma$ to be $1.92 \pm 0.01$ with $f_{pl} = 0.87^{+0.02}_{-0.01}$.

In the next observation in 2010 (X2) with XMM-Newton, we found that the accretion rate further decreased to $-1.69^{+0.03}_{-0.02}$ with corresponding $R_{\rm cor}$ as $50^{+31}_{-20}$ $R_g$. The warm corona is characterized by $kT_e$ and $\tau$, which are found to be $0.25^{+0.03}_{-0.02}$ keV and $18 \pm 3$, respectively. Notably, this observation exhibited the highest temperature of the warm corona. For this observation, $\Gamma$ and $f_{\rm pl}$ are found to be $1.72 \pm 0.02$ and $0.82^{+0.03}_{-0.09}$, respectively.

In the 2013 (S2) and 2022 (S3+NU) observations, as the soft excess was not detected in the spectral fitting, it is difficult to fit the spectra using the `Optxagnf` model. During fitting, most of the model parameters were insensitive. Therefore, we could not constrain them to a significant limit. We started our fitting by fixing the warm corona temperature at 0.10 keV for both observations. It is important to note that this is the lower limit of $kT_e$, and the model is insensitive to this parameter. We also calculated the upper limit of the optical depth and coronal radius for these observations. Additionally, the fraction of energy emitted as the power-law component ($f_{\rm pl}$) was constrained to an upper limit of 0.95 for both observations. The photon indices for these observations are found to be $1.79^{+0.04}_{-0.03}$ and $1.76^{+0.02}_{-0.04}$, respectively. The values of the best-fitted parameter are quoted in Table 5. We show in Figure 4 the results of the MCMC analysis for the best-fit `Optxagnf` model parameters found from the S1 and X1, and the best-fitting spectrum along with the residuals presented in Figure 5.

### 3.4.2. Relativistic Reflection

Another possible origin of the soft X-ray excess is the relativistically blurred reflection from a photoionized accretion disk (J. A. Garcìa et al. 2019; R. Ghosh & S. Laha 2020; G. A. Matzeu et al. 2020; Y. Xu et al. 2021; R. Ghosh et al. 2022; N. Kumari et al. 2023; Z. Yu et al. 2023). When emission from the primary continuum originating from the corona or Compton cloud illuminates the colder accretion disk, a reflection spectrum with fluorescence lines and other spectral features is produced (R. R. Ross & A. C. Fabian 2005). These emission lines are then blurred and distorted by relativistic effects (A. Laor 1991; J. Crummy et al. 2006) as they originate close enough to the supermassive black hole. This process can generate a smooth spectrum below 2 keV, commonly called soft excess.

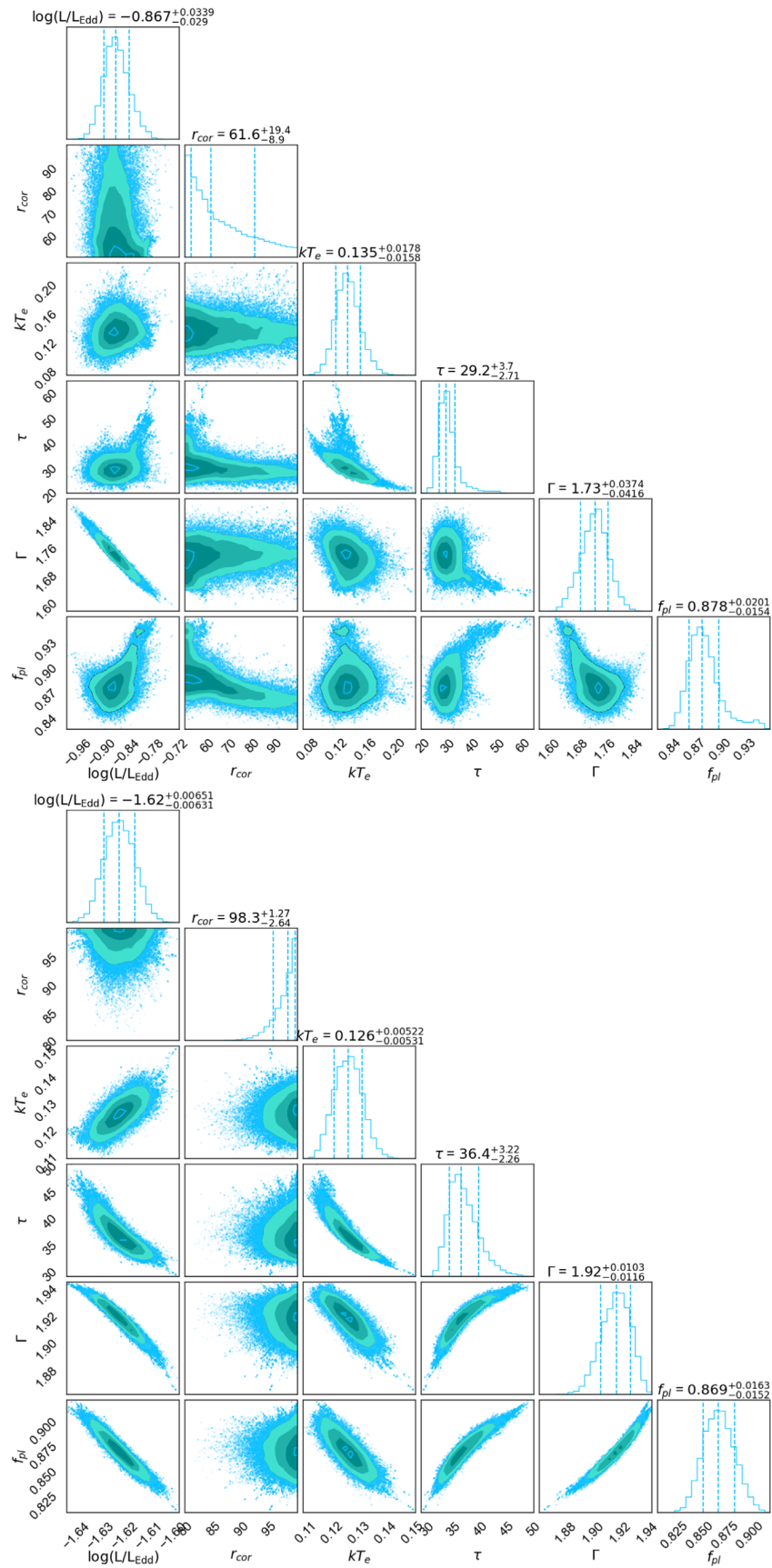

**Figure 4.** Corner plots of spectral parameters from MCMC analysis for the S1 (top) and X1 (bottom) using `Optxagnf` model. One-dimensional histograms represent the probability distribution. Three vertical lines in the 1D distribution show 16%, 50%, and 90% quantiles. We used CORNER.PY (D. Foreman-Mackey 2017) to plot the distributions. The units of $R_{cor}$ and $kT_e$ are $R_g$ and keV, respectively.

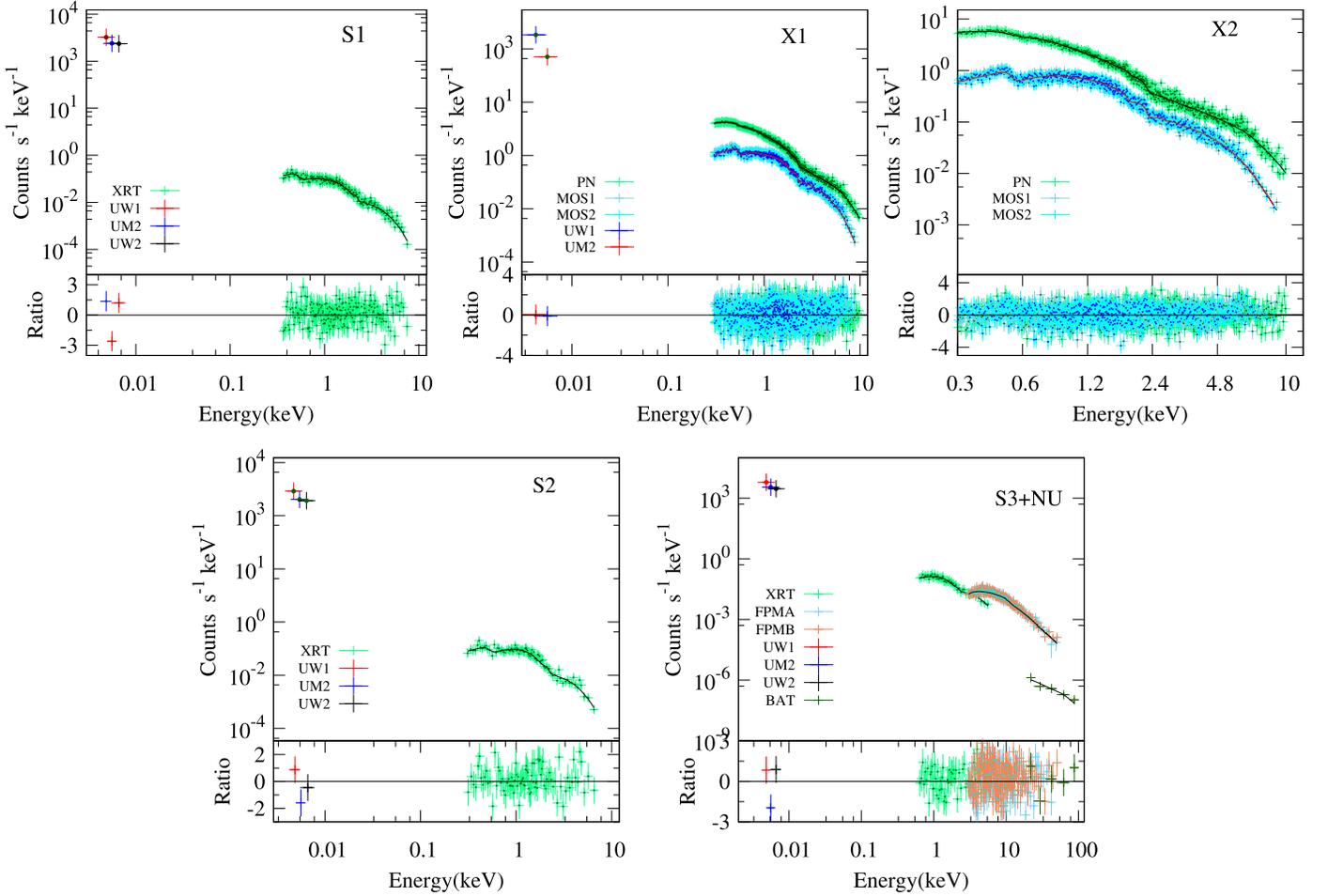

**Figure 5.** Optxagnf model fitted spectra of Mrk 50 from the different epochs of observations along with the residuals obtained from the spectral fitting.

To investigate the presence of relativistic reflection in the spectra of Mrk 50, we adopted the Relxillcp model (J. A. Garcìa et al. 2018), a variant of the relativistic reflection model Relxill (J. A. Garcìa et al. 2013, 2014; T. Dauser et al. 2014; T. Dauser et al. 2016). The Relxillcp model uses the nthcomp model (A. A. Zdziarski et al. 1996; P. T. Życki et al. 1999) as a simple Comptonization model to calculate the primary source spectrum. This variant does not assume any particular geometry, and the primary continuum emission is characterized by the spectral index ($\Gamma$) and the temperature of the corona ($kT_e$).

The reflection fraction ($R_f$), a parameter of the Relxillcp model, is defined as the ratio between the Comptonized emission directed toward the disk and that escaping to infinity. The emission profile is modeled as a broken power law with $E(r) \sim r^{-q_1}$ for $r > R_{br}$ and $E(r) \sim r^{-q_2}$ for $r < R_{br}$, where $E(r)$, $q_1$, $q_2$, and $R_{br}$ represent the emissivity, outer emissivity index, inner emissivity index, and break radius, respectively. This model also provides information on parameters such as the ionization parameter ($\xi$), iron abundance ($A_{Fe}$), inclination angle ($i^\circ$), the inner radius of the accretion disk ($R_{in}$), and the spin parameter of the black hole ($a$).

During spectral fitting, we initially fixed the black hole spin parameter to the maximum value (0.998) and set the inner radius ($R_{in}$) to 1.24$R_g$, the lowest value allowed in the model. In the second scenario, we fixed the spin parameter to zero and set the inner radius to 6$R_g$. The fit statistics did not improve for the rotating or nonrotating scenarios across all of our observations. The maximally spinning black hole scenario has also been observed in previous studies in other Seyfert 1 galaxies (J. A. Garcìa et al. 2019; S. G. H. Waddell et al. 2019; Y. Xu et al. 2021). Therefore, we continued with the rotating black hole scenario for spectral fitting, and we kept the spin parameter and inner radius fixed at 0.998 and 1.24$R_g$, respectively.

We start our analysis using the Relxillcp model and found that this model alone could not fit all of the observed spectra from UV to X-ray bands. While it fits the spectra above 0.3 keV for all observations, it does not account for the lower-energy UV spectra. We added a power-law component to the model to fit the broadband spectra from UV to X-ray ranges for all observations. Therefore, the model used to fit the broadband spectra from UV to X-ray is

**Table 6**
Best-fit Parameters for Observations of Mrk 50 with the Physical Model `Relxillcp`

| Models | Parameter | S1 | X1 | X2 | S2 | S3+NU |
|---|---|---|---|---|---|---|
| `Powerlaw` | $\Gamma_{PL}$ | $1.76^{+0.07}_{-0.04}$ | $1.73 \pm 0.01$ | ... | $1.76^{+0.12}_{-0.16}$ | $1.78^{+0.14}_{-0.06}$ |
| | Norm$_{PL}^a$ | $3.52^{+0.79}_{-0.71}$ | $0.30^{+0.05}_{-0.04}$ | ... | $2.51^{+0.11}_{-0.56}$ | $1.61^{+0.21}_{-0.34}$ |
| `Relxillcp` | $q_1$ | $3.08^{+1.72}_{-0.98}$ | $3.52^{+1.00}_{-0.57}$ | $3.35^{+0.61}_{-0.43}$ | $3.02^{+1.44}_{-1.60}$ | $3.06^{+1.23}_{-0.78}$ |
| | $q_2$ | $3.00^b$ | $3.0^b$ | $3.0^b$ | $3.0^b$ | $3.0^b$ |
| | $R_{br}(R_g)$ | <47 | <100 | <80 | <58 | <32 |
| | $i$(deg) | <27 | <25 | <38 | <25 | <31 |
| | $\Gamma_{Ref}$ | $1.91^{+0.12}_{-0.04}$ | $2.78^{+0.03}_{-0.05}$ | $1.71^{+0.08}_{-0.03}$ | $1.75^{+0.06}_{-0.05}$ | $1.74^{+0.20}_{-0.26}$ |
| | $\log\xi$ | $4.03^{+0.11}_{-0.09}$ | $4.27 \pm 0.22$ | $3.48^{+0.03}_{-0.07}$ | < 3.92 | $3.28^{+0.95}_{-0.70}$ |
| | $R_f$ | $2.83^{+0.66}_{-0.21}$ | $6.53^{+1.43}_{-1.21}$ | $5.73^{+2.86}_{-1.38}$ | $0.54^{+0.72}_{-0.49}$ | $0.48^{+0.36}_{-0.24}$ |
| | Norm$_{Ref}^a$ | $2.56^{+0.32}_{-1.20}$ | $0.29^{+0.04}_{-0.05}$ | $0.05 \pm 0.02$ | $3.47^{+0.05}_{-0.03}$ | $1.36^{+0.04}_{-0.08}$ |
| | $\chi^2$/dof | 232.70/196 | 1511.71/1592 | 1790.34/1718 | 67.64/76 | 322.42/275 |

**Notes.** Spectral fitting of all observations includes simultaneous optical-UV data except X2.
[a] In units of $10^{-2}$ photons keV$^{-1}$ cm$^{-2}$ s$^{-1}$.
[b] Indicates a frozen parameter.

represented as Const × Tbabs × Redden × (Relxillcp + Powerlaw). The fitted results are presented in Table 6. In our analysis, the emissivity index ($q_2$), which corresponds to the geometry of the accretion disk in the outer region ($r > R_{br}$), was fixed at 3.0. As a result, the outer portion of the disk behaves like a Newtonian accretion disk. Conversely, the emissivity index ($q_1$) for the inner part ($r < R_{br}$) of the disk was allowed to vary freely. During spectral fitting, we observed that the iron abundance value became insensitive. Additionally, power-law continuum fitting revealed the absence of an Fe-line in Mrk 50, which may indicate a low iron abundance in this source. Therefore, we fixed the iron abundance $A_{Fe}$ at its lowest value ($0.5\,A_\odot$) for all observations. From the 2007 observation (S1), we found that the photon index from the reflection model ($\Gamma_{Ref}$) is $1.91^{+0.12}_{-0.04}$, which is marginally higher than the photon index from the power-law model, $\Gamma_{PL} = 1.76^{+0.07}_{-0.04}$. This may indicate a marginal presence of soft excess caused by reflection in the observed spectrum. We also observed that the emissivity index for the inner disk is $q_1 = 3.08^{+1.72}_{-0.98}$, which is comparable to the outer part of the accretion disk. This suggests that the entire disk follows Newtonian geometry during this observation period. From the model fitting, we determined the ionization parameter for the relativistic reflection component as $\log\xi = 4.03^{+0.11}_{-0.09}$, with a reflection fraction of $R_f = 2.83^{+0.66}_{-0.21}$. The upper limit of the break radius and the inclination angle are found to be $R_{br} < 47 R_g$ and $i < 27$ deg, respectively.

In the next observation (X1), where a strong presence of soft excess was observed, we found a substantial difference between the indices. We observed that $\Gamma_{PL} = 1.73 \pm 0.01$ and $\Gamma_{Ref} = 2.78^{+0.03}_{-0.05}$. This difference could be explained by the reflection fraction $R_f$ at $6.53^{+1.43}_{-1.21}$, which is substantially higher compared to the previous observations. We found that $q_1 = 3.52^{+1.00}_{-0.57}$, which is marginally higher than $q_2$ with high uncertainties. This may indicate that due to a higher reflection component, the inner part of the disk deviates from Newtonian geometry. Due to high uncertainty, we are unable to constrain $R_{br}$, which is found at $R_{br} < 100 R_g$. We also calculated the inclination angle for this observation through the spectral fitting with the composite model and found it to be $i < 25$ deg.

In the case of the X2 observation, the `RelxillCp` model is sufficiently accurate to fit the observed spectrum. Therefore, we excluded the `Powerlaw` component from our baseline model for this observation. Since UV observations are unavailable, we infer that the power-law component is dominant in the UV domain, while the `RelxillCp` model is used to fit the X-ray component of the observed spectrum. A soft excess was detected in this observation, leading to a higher reflection coefficient ($R_f = 5.73^{+2.86}_{-1.38}$) with respect to other observations. The inner part of the disk also marginally deviated from the Newtonian approach by considering $q_1 = 3.35^{+0.61}_{-0.43}$ with respect to $q_2 = 3.0$. The photon index was found to be at $\Gamma_{Ref} = 1.71^{+0.08}_{-0.03}$ with an ionization parameter $\log\xi$ at $3.48^{+0.03}_{-0.07}$. The upper limit of the break radius was determined to be $80 R_g$, and the inclination angle to be $i < 38$ deg.

For the observations S2 and S3+NU, we found that the entire disk followed Newtonian geometry at the time of the observations. The emissivity indices for the inner part of the disk are found to be $3.02^{+1.44}_{-1.60}$ and $3.06^{+1.23}_{-0.78}$ for S2 and S3+NU, respectively, which are very close to $q_2$ at 3. As the inner and outer parts of the disk are indistinguishable, we cannot draw a clear boundary between them. We found that the break radius at $58 R_g$ and $32 R_g$ represents the upper limit of this parameter for the S2 and S3+NU observations, respectively. The photon indices for the reflection model ($\Gamma_{Ref}$) are $1.75 \pm 0.6$ and $1.74^{+0.20}_{-0.26}$ for these observations, respectively. These values agree well with the photon indices from the power-law model, $1.76^{+0.12}_{-0.16}$ and $1.78^{+0.14}_{-0.06}$, respectively. As the soft excess component is absent in these spectra, the reflection coefficients for these observations are minimized. The values of this parameter are $0.54^{+0.72}_{-0.49}$ and $0.48^{+0.36}_{-0.24}$, respectively. The ionization parameters, $\log\xi$, calculated from spectral fitting are 3.92 and $3.28^{+0.95}_{-0.70}$ for the observations S2 and S3+NU, respectively. The upper limit of the inclination angles is determined as $i < 25°$ and $i < 31°$, respectively. Based on overall spectral analysis from 2007–2022, the upper limit on the inclination of the source is estimated to be $<38°$, which is consistent with the Seyfert type 1 classification criteria.

## 4. Discussion

Mrk 50 has yet to be explored in detail to date. Our main motivation is understanding the physical processes in the high-energy (UV/X-ray) regime around the central supermassive black hole. Along with this, we also explored the physical origin of the soft excess component and its variability. We

**Table 7**
Fluxes of the Different Spectral Components of Mrk 50 Obtained from the Observations Used in Our Work

| Spectral Component | Flux S1 | Flux X1 | Flux X2 | Flux S2 | Flux S3+NU |
|---|---|---|---|---|---|
| Soft X-ray Excess[a] ($\times 10^{-12}$) | $1.86^{+0.46}_{-0.85}$ | $2.75^{+0.27}_{-0.30}$ | $1.31^{+0.15}_{-0.21}$ | <0.03 | <0.04 |
| Hard X-ray Flux ($\times 10^{-12}$) | $8.51^{+0.55}_{-0.56}$ | $9.12^{+0.21}_{-0.21}$ | $7.07^{+0.12}_{-0.12}$ | $7.76^{+0.45}_{-0.77}$ | $6.91^{+0.12}_{-0.13}$ |
| Total X-ray Flux ($\times 10^{-11}$) | $2.04^{+0.08}_{-0.05}$ | $2.85^{+0.03}_{-0.01}$ | $1.76^{+0.02}_{-0.03}$ | $1.77^{+0.08}_{-0.09}$ | $1.12^{+0.03}_{-0.03}$ |
| UVmonochromatic[b] | | | | | |
| UVW2 ($\times 10^{-15}$) | $10.94 \pm 0.20$ | ⋯ | ⋯ | $7.80 \pm 0.18$ | $4.49 \pm 0.11$ |
| UVM2 ($\times 10^{-15}$) | $8.77 \pm 0.19$ | $7.26 \pm 0.07$ | ⋯ | $6.63 \pm 0.18$ | $4.10 \pm 0.14$ |
| UVW1 ($\times 10^{-15}$) | $7.23 \pm 0.14$ | $6.23 \pm 0.10$ | ⋯ | $6.10 \pm 0.15$ | $3.80 \pm 0.11$ |
| $F_{2\mathrm{keV}}^{a}$ ($\times 10^{-12}$) | $2.70^{+0.09}_{-0.11}$ | $1.82^{+0.04}_{-0.05}$ | $1.18^{+0.03}_{-0.02}$ | $1.28^{+0.07}_{-0.24}$ | $1.13^{+0.02}_{-0.03}$ |
| $\alpha_{\mathrm{OX}}$ | $1.20 \pm 0.01$ | $1.24 \pm 0.01$ | ⋯ | $1.29 \pm 0.02$ | $1.24 \pm 0.01$ |

**Notes.**
[a] The X-ray fluxes are in units of erg cm$^{-2}$ s$^{-1}$.
[b] The UV monochromatic fluxes are measured from Swift UVOT and XMM-Newton OM instruments. The fluxes are in units of erg cm$^{-2}$ s$^{-1}$ Å$^{-1}$.

performed a detailed temporal and spectral study of Mrk 50 based on long-term observations from 2007–2022, using data from various X-ray observatories such as Swift, XMM-Newton, and NuSTAR. From the temporal analysis, we found that Mrk 50 showed <10% variability in different energy bands, except during the 2007 (S1) observation, where the source variability was around 20% in the X-ray domain. From 2013 onward, the source became nonvariable ($\sqrt{\text{variance of rate}}$ < = mean error of the rate).

### 4.1. Mrk 50 as a "Bare" AGN

In our spectral analysis of Mrk 50 (Section 3.2), we found that the emission in the low-energy domain, starting from 1 eV to ∼3 keV, is unaffected by the intrinsic hydrogen column densities along the line of sight. We used various absorption models to detect the contribution of neutral or ionized hydrogen along the line of sight to the observed spectra. However, each model failed to detect any trace of extragalactic hydrogen along the line of sight. Therefore, all observations of Mrk 50 from 2007–2022 are free from neutral and/or ionized absorptions in the UV/X-ray domain. Similar results are reported for other AGNs with a "bare" nucleus at the center (S. Vaughan et al. 2004; D. J. Walton et al. 2010; P. Nandi et al. 2023, 2024). The lack of intrinsic absorption has also been previously reported for Mrk 50 by R. V. Vasudevan et al. (2013). Our preliminary examination of these observations confirms that Mrk 50 is indeed a "bare" Seyfert 1 nucleus during the observational period from 2007–2022. For further clarification, one would need to examine optical observations, which is beyond the scope of this work.

### 4.2. Long-term Spectral Variability

In the previous Sections, we presented spectral fit results using several phenomenological and physical models applied to the data of Mrk 50. We observed that in the S1 observation, the total X-ray flux (0.3–10 keV) was $2.04^{+0.08}_{-0.05} \times 10^{-11}$ erg cm$^{-2}$ s$^{-1}$, which increased by approximately 1.4 times (∼2.85 ± 0.03 × $10^{-11}$ erg cm$^{-2}$ s$^{-1}$) in the X1 observation. However, within a year, the flux dropped by a factor of approximately 1.5 in the X2 observation. The calculated flux in X2 observation was $1.76^{+0.02}_{-0.03} \times 10^{-11}$ erg cm$^{-2}$ s$^{-1}$. After that, the total X-ray flux decreased to 1.12 ± 0.03 × $10^{-11}$ erg cm$^{-2}$ s$^{-1}$ in the most recent observation S3+NU, which is approximately 2.5 times less than the flux in the X1 observation. From Table 7, it is evident that both the hard X-ray and UV monochromatic flux varied significantly between the observations (see Figure 6). The UV monochromatic flux (UVW1) was 7.23 ± 0.14 × $10^{-15}$ erg cm$^{-2}$ s$^{-1}$ Å$^{-1}$ in S1 and then declines to 3.80 ± 0.11 × $10^{-15}$ erg cm$^{-2}$ s$^{-1}$ Å$^{-1}$ in S3+NU observation. The variation in the X-ray flux over time may arise due to changes in the strength of Comptonization inside the Compton cloud. It is also believed that these variations may be attributed to changes in the UV photon flux. Simultaneous UV and X-ray observations allow us to probe these types of spectral variability. For this purpose, a hypothetical power law between 2500 Å and 2 keV, known as $\alpha_{\mathrm{OX}}$ (H. Tananbaum et al. 1979), is estimated from the simultaneous UV and X-ray observations. The parameter $\alpha_{\mathrm{OX}}$ is defined as $\alpha_{\mathrm{OX}} = -0.384 \log[L_{2500}/L_{2\mathrm{keV}}]$. We used the UVW1 band to compute $\alpha_{\mathrm{OX}}$ because it has the closest effective wavelength to 2500 Å. The observed $\alpha_{\mathrm{OX}}$ values remain nearly constant at ∼1.24 across all observations, suggesting that the X-ray and UV photons vary in a similar manner. The steeper value of this parameter indicates that the X-ray emission is dimmed more than the UV emission. This phenomenon is commonly observed in the case of most AGNs (I. V. Strateva et al. 2005). According to L. C. Gallo (2006), steep $\alpha_{OX}$ is often accompanied by X-ray spectral complexity such as enhanced blurred reflection. Due to relativistic effects, most of the emission from the primary power-law component bends toward the black hole, preventing it from reaching the observer (G. Miniutti & A. C. Fabian 2004; M. Pal et al. 2018). In comparison, the UV emission (produced by the colder accretion disk) remains relatively unchanged. A relatively weaker X-ray and constant UV emission makes $\alpha_{OX}$ steep.

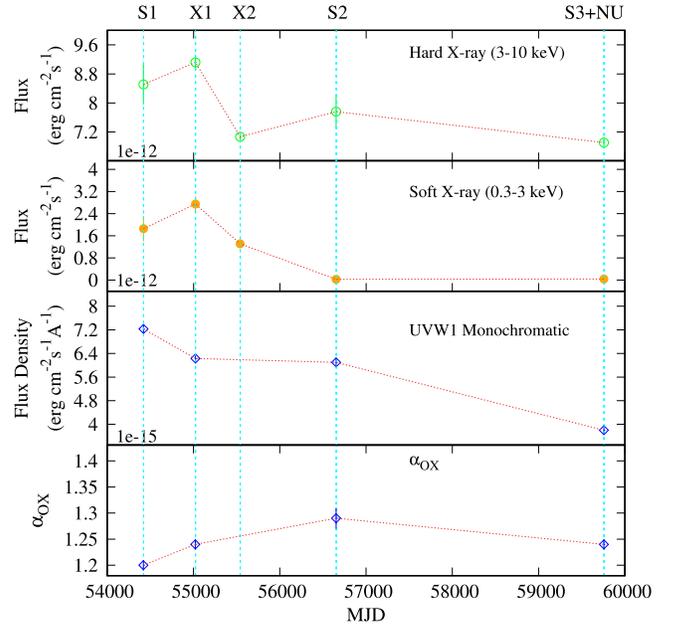

**Figure 6.** Temporal variation of soft X-ray excess (0.3–3 keV), hard X-ray (3–10 keV), UV monochromatic (UVW1) flux, and $\alpha_{ox}$ from all of the observations used in the present work are plotted.

During the spectral fitting, the softest spectrum is observed in the S1 observation with a primary continuum power-law slope of $1.88^{+0.27}_{-0.28}$ (see Table 3). The normalized accretion rate for this observation is estimated to be $-0.87 \pm 0.03$ (see Table 5), the highest among all observations. As the accretion rate is high, the emission from the disk is significant, indicating an increased supply of seed photons. As a result, a soft spectrum is observed for this observation. With the increase in the supply of seed photons, more photons interact within the Compton cloud, increasing the luminosity of the primary continuum. For this observation, the luminosity of the primary continuum is found to be $\log L_{PC} = 43.58^{+0.02}_{-0.04}$, the highest among all of the observations used in the present work. Due to the high accretion rate, the UV flux also increased. We observed the highest UV fluxes of $10.94 \pm 0.20 \times 10^{-15}$, $8.77 \pm 0.19 \times 10^{-15}$, and $7.23 \pm 0.14 \times 10^{-15}$ erg cm$^{-2}$ s$^{-1}$ Å$^{-1}$ (see Table 7) for the wavelength 1928 Å(UVW2), 2310 Å(UVM2), and 2930 Å(UVW1), respectively, for this observation. As both UV and X-ray fluxes increased, we found the UV to X-ray slope, $\alpha_{OX} = 1.20$ (see Table 7), which is comparable with other observations.

From the spectral analysis, a marginal presence of soft excess was observed in the S1 observation (see the first panel of Figure 2). Initially, this excess component was characterized by a power law with a photon index ($\Gamma_{SE}$) of $1.91^{+0.03}_{-0.04}$ (see Table 3). Later, it was described by a multicolor disk blackbody (Diskbb) with inner disk temperature ($kT_e$) at $0.16^{+0.02}_{-0.01}$ keV (see Table 4). If the soft excess is attributed to the Comptonization of seed photons from a warm corona, the optical depth of the corona is estimated to be $29^{+4}_{-3}$, with a dimension of $62^{+19}_{-9}$ $R_g$ (see Table 5). On the other hand, if the soft excess results from reflection processes, the reflection coefficient is determined to be $R_f = 2.83^{+0.66}_{-0.21}$ with a photon index $\Gamma_{Ref} = 1.91^{+0.12}_{-0.04}$ (see Table 6).

After a 2 yr gap from the S1 observation, the next X-ray observation was conducted with XMM-Newton in 2009 (X1). The X-ray continuum became hard during this observation with a photon index of $\Gamma_{PC} = 1.72 \pm 0.05$. The luminosity of the primary continuum marginally decreased from $\log L_{PC} = 43.58^{+0.02}_{-0.04}$ to $\log L_{PC} = 43.36^{+0.02}_{-0.01}$ (see Table 3) within 2 yr. A decrease in luminosity suggests a decrease in the accretion rate, estimated to be $-1.62 \pm 0.01$, lower than the previous observation (see Table 5).

With decreasing accretion rate, the number of seed photons decreases. The decreasing number of seed photons is unable to cool the warm corona efficiently due to less scattering and is most likely responsible for the expansion of the warm corona. In this work, we found an increase in the value of $r_{cor}$ and a decrease in the accretion rate as found in other AGNs, e.g., Mrk 1018 (H. Noda & C. Done 2018) and NGC 1566 (P. Tripathi & G. C. Dewangan 2022). This decrease in accretion rate also results in a reduction in UV flux, measured at $7.26 \pm 0.07 \times 10^{-15}$ and $6.23 \pm 0.10 \times 10^{-15}$ erg cm$^{-2}$ s$^{-1}$ Å$^{-1}$ for wavelengths 2310 Å and 2930 Å, respectively (see Table 7). Although both UV and X-ray fluxes decrease, the slope remains constant ($\alpha_{OX} = 1.20$, as noted in Table 7).

From spectral analysis, a strong soft excess is observed in this observation (second panel of Figure 2). This soft excess is characterized by a power law with photon index and luminosity of $\Gamma_{SE} = 2.63 \pm 0.04$ and $\log L_{SE} = 43.64 \pm 0.01$, respectively. This represents the highest soft excess luminosity among all observations from 2007 (S1) to 2022 (S3+NU). The optical depth and coronal radius are also found to be at maximum ($\tau = 36^{+3}_{-2}$ and $r_{cor} = 98^{+1}_{-3}$ $R_g$; see Table 5) during this observation. This higher optical depth and larger corona generate more soft X-ray photons, resulting in the strong presence of soft excess. The reflection model suggests a stronger reflection during this observation, with the inner disk deviating from Newtonian geometry (inner emissivity index $q_1 = 3.52^{+1.00}_{-0.57}$), producing a steeper power law with photon index $\Gamma_{Ref} = 2.78 \pm 0.04$.

During the X2 observation, the observed soft excess is relatively low, irrespective of the similar nature of the primary power-law continuum. The slope of the primary continuum and corresponding luminosity during this observation are $\Gamma_{PC} = 1.70 \pm 0.05$ and $\log L_{PC} = 43.38^{+0.03}_{-0.02}$, respectively, consistent with the X1 observation. The soft excess luminosity declined from $\log L_{SE} = 43.64^{+0.02}_{-0.01}$ during X1 observation, to $\log L_{SE} = 43.15^{+0.02}_{-0.01}$ during X2 observation. This decrease in the soft excess can be explained by the warm Comptonization model. Using this model, we found that the optical depth decreased from $\tau = 36^{+3}_{-2}$ to $\tau = 18 \pm 2$, while the size of the warm corona contracted from $98^{+1}_{-3}$ $R_g$ to $50^{+31}_{-20}$ $R_g$. A smaller size of warm corona with a relatively small optical depth produces a smaller number of X-ray photons in the soft X-ray band. Therefore, the observed decrease in the soft excess luminosity during observation X2, compared to that during X1, is due to the decrease in the size and optical depth of the warm corona. We also attempted to describe the soft excess using the reflection model. In this model, we also noticed that the value of the reflection coefficient decreased from $6.53^{+1.43}_{-1.21}$ to $5.73^{+2.86}_{-1.38}$. As a result, fewer photons were effectively reflected to produce excess emission below 2 keV.

During the S2 observation, we encountered the hardest spectrum of Mrk 50 with an index and corresponding luminosity of the primary continuum as $\Gamma_{PC} = 1.54^{+0.42}_{-0.40}$, and $\log L_{PC} = 43.46^{+0.21}_{-0.36}$, respectively. From the physical model fitting, the accretion rate is estimated to be $(-1.77^{+0.05}_{-0.03})$, the lowest among all of the observations. We also noticed that the amount of soft excess is remarkably reduced and falls below the detection limit. This is possibly due to the low exposure time of this observation. In this case, the slope of the power-law representing the soft excess is comparable with the slope of the primary continuum. Although the effective size of the warm corona was not reduced at the time of observation, the optical depth decreased from $\tau = 18 \pm 3$ (X2 observation) to $\tau < 5$ (S1 observation). Consequently, very few excess photons are produced in the soft X-ray range. Therefore, most of the photons observed in the soft X-ray regime are from the primary continuum through the process of Comptonization in the hot corona. From the reflection model perspective, we found that the reflection coefficient is $0.52^{+0.72}_{-0.49}$, much lower than other observations. As a result, fewer reflected photons are produced to contribute toward the excess emission in the soft X-ray domain. Additionally, we found that the dimensions of the Compton cloud or corona are of the same order as calculated from different models for this observation.

The latest observation in the X-ray domain was conducted with the Swift and NuSTAR observatories in 2022 (S3+NU). Using these observations, we obtained a broadband spectrum of Mrk 50 starting from UV to hard X-ray range. We noticed that while extending the primary continuum up to 100 keV, no deviation was seen in the spectra, indicating the absence of a reflection hump above 10 keV. During this observation, the primary continuum luminosity was $\log L_{PC} = 43.14^{+0.02}_{-0.06}$, the lowest among all of the observations, with $\Gamma_{PC} = 1.77 \pm 0.07$. The normalized accretion rate increased from $-1.77^{+0.05}_{-0.03}$ (S2 observation) to $-1.20^{+0.03}_{-0.05}$ during this observation. Since the soft excess component is below the detection limit, determining parameters associated with the warm corona is challenging. However, spectral analysis indicates an optical depth of $\tau < 3$ and a transitional radius of $r_{cor} > 10 R_g$ for this observation. Therefore, the observed broadband spectrum (from UV to hard X-ray range) is dominated by the primary continuum with power-law fraction $f_{pl} > 0.95$. From the reflection model perspective, the whole disk followed the Newtonian geometry with a negligible amount of reflection ($R_f = 0.48^{+0.36}_{-0.24}$). Thus, photon contribution in the soft X-ray band is nearly negligible. Consequently, a single power-law model is sufficient to fit the broadband spectrum from 1 eV to 100 keV.

For the overall picture, we observed that the slope of the primary continuum varied from $1.54^{+0.42}_{-0.40}$ to $1.88^{+0.27}_{-0.28}$. However, the primary continuum luminosity remained nearly constant, with an average value of $\log L_{PC}^{avg} = 43.38^{+0.19}_{-0.18}$. In contrast, for the soft excess component (below 3 keV), we found that both the power-law index ($\Gamma_{SE}$) and luminosity ($\log L_{SE}$) varied within the ranges of $1.77$–$2.63 \pm 0.04$ and $42.91$–$43.64^{+0.02}_{-0.01}$, respectively. We explored the possible physical reasons for this spectral variability and found that the variation in the primary continuum can be explained in terms of accretion dynamics. Depending on the accretion rates and the efficiency of the Compton cloud, the slope of different spectral components and the corresponding luminosity change. The long-term variations of a few spectral parameters are presented in Figure 7.

### 4.3. Relation between Different Spectral Parameters

In the previous Section, we explained the observed spectral variabilities in Mrk 50 using various spectral parameters derived from different model fittings (see Section 3.2). Although all of the model parameters contribute to the overall spectral variability, they are not entirely independent of each other. In this subsection, we examine the correlations between different spectral parameters. We calculated the correlation coefficient[18] (CC) to calculate the degree of correlation between the parameters. The correlations between some selected parameters are shown in Figure 8. This limited number of observations (five) in a period of 15 yr makes it challenging to draw definitive physical conclusions about the correlations between different parameters. However, we discuss the observed trends in the correlation between different parameters on this 15 yr timescale. Based on these correlation studies, we infer the nature of the source and understand the overall spectral properties of Mrk 50.

---
[18] https://www.statskingdom.com/correlation-calculator.html

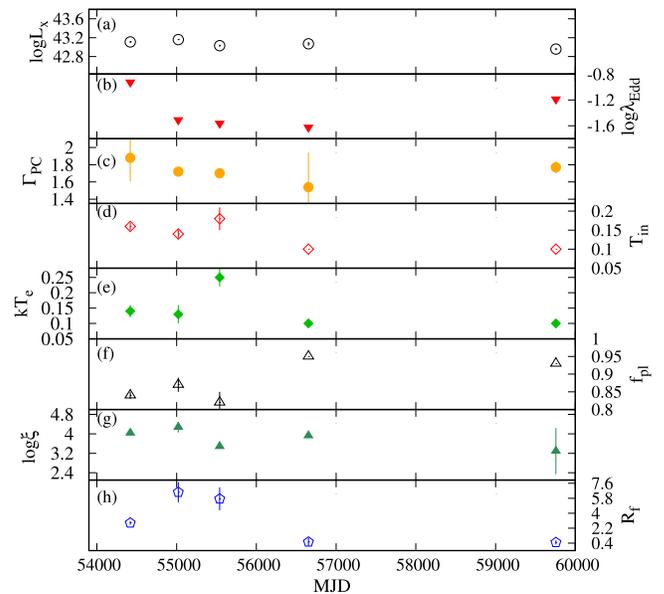

**Figure 7.** Temporal variation of different parameters obtained from the phenomenological and physical model fittings to the observed spectra of Mrk 50.

We began our spectral analysis by parameterizing the observed spectra using a power-law model, calculating the spectral slopes and luminosities and/or fluxes of different spectral components, including the soft excess and the primary continuum. We observed hints of correlations between luminosities and fluxes in the hard and soft X-ray regions, with CCs of 0.66 ($p$-value $= 0.22$) and 0.78 ($p$-value $= 0.11$), respectively, as shown in Figure 8 (panels (a) and (b)). Similar correlations have been observed in other AGNs, such as IC 4329A (P. Tripathi et al. 2021), PKS 0558-504 (M. Gliozzi et al. 2013), and NGC 7469 (K. Nandra & I. E. Papadakis 2001) and have been interpreted in terms of thermal Componization of soft photons in the hot corona. As the number of soft photons increases, the amount of scattering in the hot corona also increases, leading to an enhancement in the Comptonized X-ray flux. This common feature is also observed in "bare" AGNs (P. Nandi et al. 2021, 2023).

Additionally, we found that the UV flux correlates with both soft and hard X-ray fluxes, with CC = 0.61 ($p$-value $= 0.39$) for soft X-ray versus UV, and CC = 0.79 ($p$-value $= 0.20$) for hard X-ray versus UV. These correlations indicate a relationship between the disk and the hot corona, and a link between the disk and the medium producing the soft excess. Furthermore, we observed a correlation trend between the spectral slope of the primary continuum ($\Gamma_{PC}$) and the normalized accretion rate ($\lambda_{Edd}$), with a CC of 0.89 ($p$-value $= 0.09$). The $\Gamma_{PC}$-$\lambda_{Edd}$ correlation indicates the connection between the accretion disk and the X-ray corona. The physics behind this correlation is not clear. One possible explanation is that a higher accretion rate increases the supply of soft photons; an increase in the supply of soft photons can produce more hard photons by interacting with the Compton cloud. Consequently, the power-law index steepens, leading to the observed correlation between $\Gamma_{PC}$ and $\lambda_{Edd}$ (A. C. Fabian et al. 2015; C. Ricci et al. 2018; R. V. Vasudevan & A. C. Fabian 2007; N. Layek et al. 2024).

Next, we applied physical models to the observed spectra to explain the nature and origin of the different spectral components. We found a hint of a correlation between the reflection coefficient ($R_f$) and the slope of the soft excess component ($\Gamma_{SE}$). Since the *relxillcp* model fits the X-ray spectra well, this correlation may suggest a link between $R_f$ and $\Gamma_{SE}$ in the X-ray band, implying that the soft excess component may be associated with the reflection phenomena. We also explored the correlation between the UV to X-ray index, $\alpha_{OX}$, and various spectral parameters, finding that $\alpha_{OX}$ is anticorrelated with $\lambda_{Edd}$, with CC = −0.87 (*p*-value = 0.11). Since $\alpha_{OX}$ remains nearly constant at 1.24 across our observations, and the values from different observations fall within the range of uncertainties, it is challenging to comment on this anticorrelation conclusively. The dependence of $\alpha_{OX}$ with $\lambda_{Edd}$ suggests that the disk/corona relative intensity also depends on the accretion rate (R. Fanali et al. 2013).

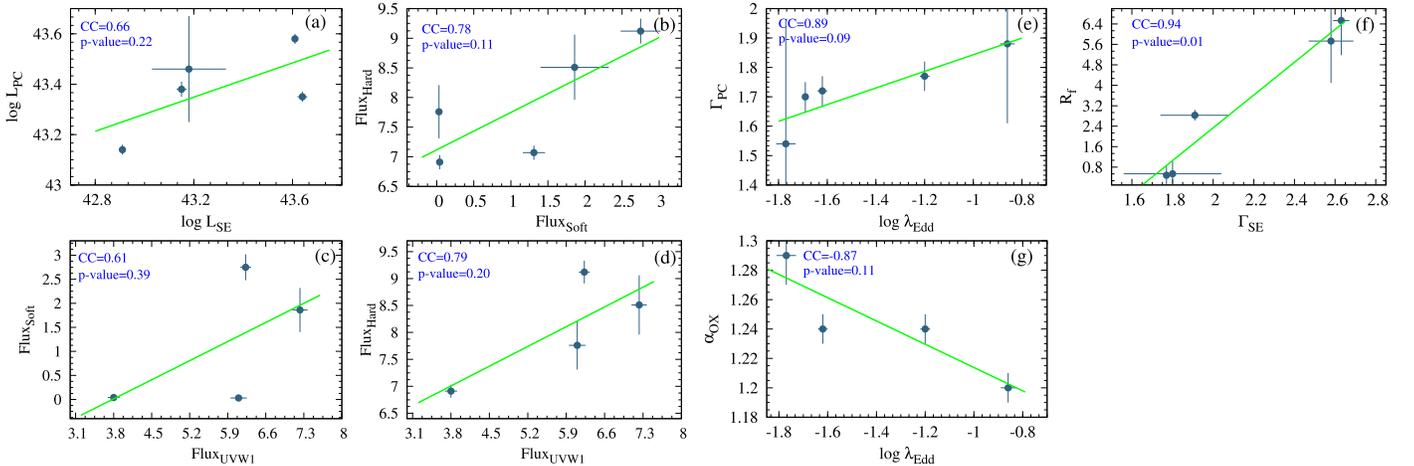

**Figure 8.** Correlation between different spectral parameters.

In this study, we explored correlations between various parameters and found hints of correlations between the spectral parameters. These correlations can be explained from a physical perspective. However, it is important to note that the *p*-values associated with these correlations are >0.05, indicating that the correlations may not be statistically significant at the 5% level. This suggests that the observed relationships could potentially be due to random variability. Further investigation or a larger sample size may be needed to draw more accurate conclusions.

### 4.4. Soft Excess

An enhancement of flux in the soft X-ray range (below ∼2 keV) over the primary power-law continuum is commonly observed in most of the Seyfert 1 AGNs. This excess in the low-energy range is known as soft excess. Our spectral analysis revealed the presence of a prominent soft excess below 2 keV in Mrk 50. However, this soft excess component is found to change over time (see Figure 9). From the flux calculations, we observed significant variability in soft excess flux between observations of Mrk 50 (see Table 7). This appearance and disappearance of the soft excess component intrigued us to explore its origin. Although the origin of this excess emission in the soft X-ray domain is a long-standing and unsolved puzzle in AGN studies, we attempted to unravel this mystery using various models discussed in Section 3.3.

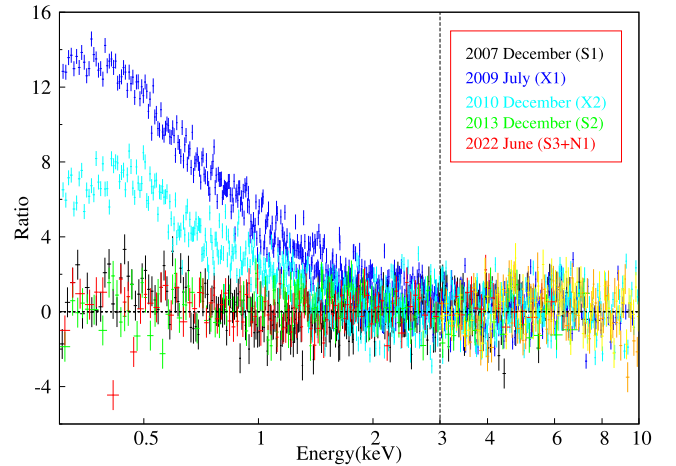

**Figure 9.** Ratio plot of the observed spectra over a power-law model, fitted to the 3–10 keV band and extrapolated to lower energies.

#### 4.4.1. Strength of Soft Excess

Among all of the observations used in the present work, the observed soft excess component is the strongest in the 2009 observation (X1). During this observation, the Fe $K_\alpha$ line and the Compton hump are also not detected in the spectrum. Spectral modeling in such cases is often challenging due to the limited number of spectral components. Additionally, the UV/X-ray combined spectra were not explored. Therefore, we initially characterized the soft excess component with a simple power law over the primary continuum.

Although the initial characterization of the soft excess component using a power-law model provided valuable insights into the temporal variation of this component, the variation of its strength relative to the primary continuum still needs to be explored. We employed a multicolored blackbody model Diskbb (K. Mitsuda et al. 1984), replacing the power-law component for the soft excess. We fitted all observed spectra with this model and found that the blackbody temperature varies within a narrow range from 0.10–0.22 keV. The soft excess is modeled as a blackbody, and its strength ($S_{SE}$) is defined as the ratio of the luminosity in the blackbody component to the luminosity in the power law between 3 and 10 keV, as described by R. V. Vasudevan et al. (2013). The parameter $S_{SE}$ is used to investigate if there is a positive link between the strength of the soft excess and reflection. R. V. Vasudevan et al. (2013) and R. Boissay et al. (2016) used this parameter to investigate whether reflection is the cause of the soft excess.

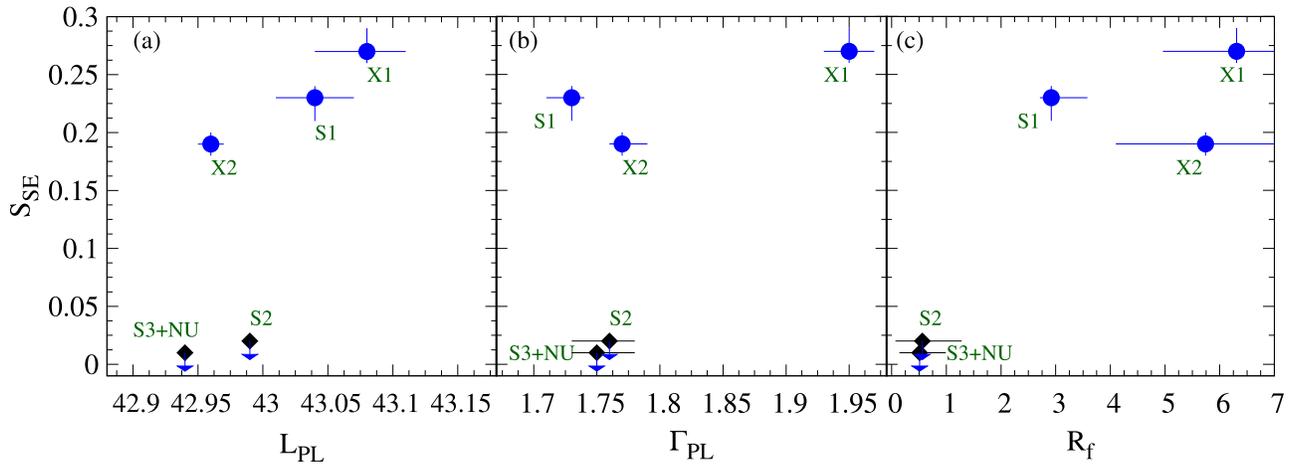

**Figure 10.** Variation of the soft-excess strength ($S_{SE}$) against the power-law luminosity ($L_{PL}$), the photon index ($\Gamma_{PL}$), and the reflection fraction($R_f$) are shown for all five epochs of observation. The soft excess is modeled using the `Diskbb` model, and strength is defined as the ratio of the luminosity of the blackbody component to the luminosity of the power-law continuum between 3 and 10 keV. The downward-pointing arrows show upper-limiting soft-excess strengths corresponding to the observations that did not show detectable soft excess (using the `Diskbb` model)

In the first three observations (S1, X1, and X2), the presence of soft excess is clearly detected, whereas it is barely detected in the spectra of the observations in 2013 (S2) and 2022 (S3+NU). We examine the correlation between the strength of soft excess ($S_{SE}$) and power-law photon index ($\Gamma_{PL}$), power-law luminosity ($L_{PL}$), and reflection strength ($R_f$; see Figure 10). Although it is challenging to determine whether a correlation or anticorrelation exists between these parameters due to the limited number of observations, we observed a hint of a correlation between $S_{SE}$ and $R_f$. Such a correlation has previously been observed in a larger sample of AGNs (R. Boissay et al. 2016; R. V. Vasudevan et al. 2013).

### 4.4.2. Origin of Soft Excess

The soft X-ray excess observed in AGNs was initially represented as a high-frequency tail of the disk emission (K. P. Singh et al. 1985; K. A. Pounds et al. 1986; K. M. Leighly 1999a; P. Magdziarz et al. 1998; K. M. Leighly 1999b). In many AGNs, this soft excess could be fitted by a blackbody model with a best-fit temperature of 0.1–0.2 keV (R. Walter & H. H. Fink 1993; B. Czerny et al. 2003; A. Jana et al. 2021); however, this temperature is significantly higher than the maximum temperatures expected in AGN accretion disk. This temperature is nearly independent of the accretion rate (see Table 5). As a result, the inner disk is unlikely to be the source of the soft excess emission. Recent studies suggest that the soft excess in Seyfert 1s can be successfully described by both the intrinsic thermal Comptonization of disk photons (C. Done et al. 2012; A. Kubota & C. Done 2018; P. O. Petrucci et al. 2018; P. O. Petrucci et al. 2020) and relativistic reflections from an ionized accretion disk (H. J. S. Ehler et al. 2018; J. A. Garcìa et al. 2019; R. Ghosh & S. Laha 2020; Z. Yu et al. 2023). So, to investigate the possible origin of the soft excess, we explored two physically motivated scenarios: the warm Comptonized disk emission model (`Optxagnf`) and the blurred reflection model (`Relxillcp`).

The `Optxagnf` model has three distinct emission components: the UV bump from the disk, the soft excess component from the warm corona, and the hard X-ray power law from the hot corona, assuming that all are powered by gravitational energy released through the process of accretion (C. Done et al. 2012). The model can potentially explain the origin and variation of the soft excess. The best-fit result using this model is presented in Section 3.4.1. The results indicate that more than 80% ($f_{pl} > 0.83$) of the power law is generated from the hot corona via the process of inverse Comptonization. Only 10%–20% of the total photons in the spectrum are involved in generating the soft excess component. This result is also cross-verified by the flux calculations in Table 7. In the `Optxagnf` model, the soft excess emission is produced due to the thermal Comptonization of the disk optical-UV photons by a warm ($kT_e \sim$ 0.1–0.2 keV) optically thick ($\tau \sim$ 10–20) corona surrounding the inner region of the disk. For AGN soft excess, the soft X-ray emitting plasma temperature is degenerate with the optical depth ($\tau$; T. J. Turner et al. 2018). In the case of Mrk 50, we also found that the warm corona temperature ($kT_e$) and optical depth $\tau$ are degenerate (see Figure 4). At a particular source luminosity or spectral state, an increase in $kT_e$ in the model is compensated for by a decrease in $\tau$ in the fit and vice versa. From spectral analysis, we found that best-fit values of the temperature of the warm corona vary between $0.13 \pm 0.01$ keV and $0.25 \pm 0.03$ keV. The strength of the soft excess varies between 0.22 and 0.05. Hence, the variation in the strength of the soft excess can be explained by the change in the size of the warm corona ($R_{cor}$) and its Comptonization efficiency. During 2009 observation (X1), we found that the warm corona extends up to the maximum value $98^{+1}_{-3}\,R_g$ with an optical depth $\tau = 36^{+3}_{-2}$ and corresponding soft X-ray flux maximized in this observation period. Before (2007) and after (2010) this observation, the values of these parameters decrease. Consequently, the strength of the soft excess drops, and the flux is reduced. However, we could not determine the exact size ($R_{cor}$), electron temperature ($kT_e$), or optical depth ($\tau$) of the Comptonizing corona when the soft excess is absent.

The blurred reflection model can also explain the origin and variation of the soft excess in Mrk 50. We begin our analysis using the `Relxillcp` model. However, the absence of key reflection features, such as a prominent Fe $K_\alpha$ line and reflection hump above 10 keV, along with deviations in the UV band, indicated that pure reflection cannot explain the observed spectra of Mrk 50 across the UV to X-ray range. The deviations in the UV bands required an additional power-law component in the spectral fitting. The variation in the soft excess strength can be explained using the boundary of the inner accretion disk ($R_{br}$) and the reflection efficiency ($R_f$). The inner disk in Mrk 50 was found to be extended up to $100 R_g$ with a maximum reflection efficiency of $6.53^{+1.43}_{-1.21}$ during the 2009 observation (X1), where the strength of the soft excess was maximum. The break radius (boundary of the inner disk) and the reflection efficiency decreased as the strength of the soft excess decreased before (S1) and after (X2, S2, and S3+NU) this observation (see Table 5).

When we consider only X-ray data, both the reflection and warm corona scenarios could explain the origin of the soft excess in Mrk 50. Both models are statistically acceptable while fitting data beyond 0.3 keV. However, the origin of the soft excess can be further investigated through timing analysis. The CCF analysis is another approach to distinguish the physical mechanisms behind the origin of soft excess in AGNs. This is a model-independent method to study the correlations and delays between the soft and hard X-ray bands. In the relativistic blurred reflection model, the soft excess is known to have originated due to the ionized reflection of hard X-ray photons from the hot corona in the inner region of the accretion disk. Therefore, one can expect a short delay (∼30 s) between the hard and soft X-ray emission, observed in narrow-line Seyfert 1 AGNs (A. C. Fabian et al. 2009; D. Emmanoulopoulos et al. 2011). On the other hand, a longer delay could be produced by Comptonization. In the context of the Optxagnf model, the soft excess emission is attributed to the disk, with the inner region acting as an optically thick Comptonizing corona (C. Done et al. 2012). An extensive study on the soft X-ray time lag reports that the lags originate from the inner part of the accretion disk, which is highly dependent on the mass of the central object (B. De Marco et al. 2013). However, our timing study shows no correlation between the soft and hard bands (see Figure 1). This suggests that both the Comptonization and reflection processes contribute toward the origin of soft excess in Mrk 50, as in the case of AGN, ESO 141-G055 (R. Ghosh & S. Laha 2020). Based on spectral analysis, we tentatively conclude that the warm Comptonization model is a favorable explanation for the soft excess in Mrk 50. However, we cannot rule out the relativistic reflection model if we consider only X-ray data beyond 0.3 keV in the spectral fitting. Detailed investigation on the origin of soft excess in Mrk 50 using different models requires high-quality simultaneous observations by XMM-Newton and NuSTAR, as well as future X-ray missions like Athena, which is beyond the scope of this work.

## 5. Conclusions

For the first time, we presented a temporal and spectral study of the Seyfert 1 AGN Mrk 50 based on long-term observations over 15 yr (2007–2022). Our key findings are as follows.

1. Based on our spectral analysis, we notice that the observed spectra are almost unaffected by intrinsic hydrogen column densities, indicating that Mrk 50 has a "bare" nucleus at its center.

2. From the ∼15 yr (2007–2022) long observations of Mrk 50, we observe that the source moved from a comparatively soft state ($\Gamma_{PC} = 1.88 \pm 0.27$) in 2007 to a comparatively hard state ($\Gamma_{PC} = 1.54 \pm 0.41$) in 2013, and then back to a soft state ($\Gamma_{PC} = 1.77 \pm 0.07$) in 2022.

3. The observed variation in the spectral slope ($\Gamma_{PC}$) can be attributed to the dynamics of the accretion flow surrounding the central supermassive black hole and the characteristics of the Compton cloud. Our analysis reveals that the normalized accretion rate decreases from $-0.87 \pm 0.03$ to $-1.77 \pm 0.05$, while $\Gamma_{PC}$ changes from $1.88 \pm 0.27$ to $1.54 \pm 0.41$. The accretion rate then increases to $-1.20 \pm 0.02$, corresponding to $\Gamma_{PC} = 1.77 \pm 0.07$. During these spectral transitions of Mrk 50, the warm corona, as described by the warm Comptonization model, and the inner disk boundary, as described by the blurred reflection model, change accordingly.

4. During our spectral analysis, the soft excess component varies throughout the observational period. In the 2009 observation, we found the strongest presence of this component in the observed spectrum. However, before and after 2009, this component gradually faded. The appearance and disappearance of the soft excess component can be explained by the dynamics of accretion and the properties of the Compton cloud.

5. The warm Comptonization model and the blurred reflection model are used to explain the spectral appearance and disappearance of the soft excess component in Mrk 50. Both models provide a satisfactory explanation for the variation in the soft excess. According to these models, the inner part of the disk plays a crucial role in generating the soft excess component, and variations in this part of the disk result in changes in the intensity of the soft excess. The soft excess appears when the warm corona is present. On the other hand, according to the reflection model, the intensity of the soft excess increases when the amount of reflection is high.

6. From the temporal properties of Mrk 50, we found that the source was variable during the 2007 observation. After that, Mrk 50 became nonvariable throughout the observational period from 2009–2022. Additionally, our cross-correlation study found no correlation between the soft and hard bands. This may indicate that both the Comptonization and reflection processes contribute toward the origin of soft excess in Mrk 50.

### Acknowledgments

We sincerely thank the anonymous referee for providing insightful comments and constructive suggestions that helped us to improve the manuscript. The research work at the Physical Research Laboratory, Ahmedabad, is funded by the Department of Space, Government of India. A.J. acknowledges support from the FONDECYT Postdoctoral fellowship (3230303). The data and the software used in this work are taken from the High Energy Astrophysics Science Archive Research Center (HEASARC), which is a service of the Astrophysics Science Division at NASA/GSFC and the High Energy Astrophysics Division of the Smithsonian Astrophysical Observatory. This work has made use of data obtained from the NuSTAR mission, a project led by Caltech, funded by NASA and managed by NASA/JPL, and has utilized the NuSTARDAS software package, jointly developed by the ASDC, Italy and Caltech, USA. This work made use of data Swift supplied by the UK Swift Science Data Centre at the University of Leicester. This research has used observations obtained with XMM-Newton, an ESA science mission with instruments and contributions directly funded by ESA Member States and NASA.

*Facilities*: HEASARC, NuSTAR, XMM-Newton, Swift.

*Software:* HEASOFT (v6.30), XSPEC (v-12.12.1, K. A. Arnaud 1996)

## Appendix

The values of the cross-calibration factors are given in Table A1 and the light curves at different epochs of observation shown in Figure A1.

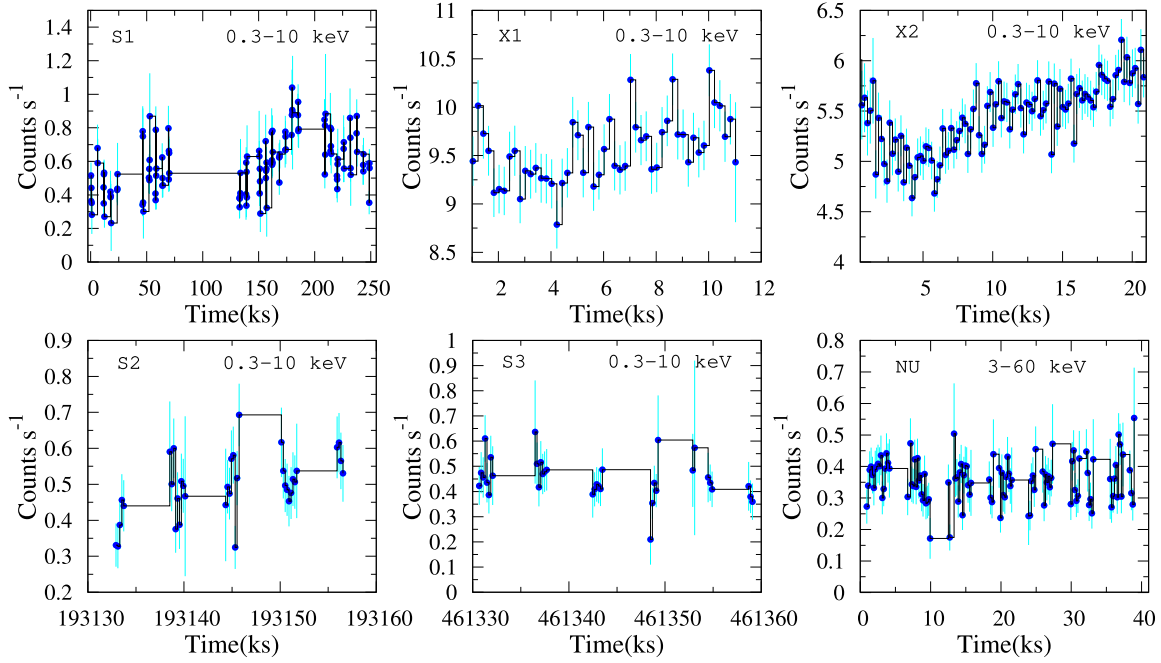

**Figure A1.** Variation of photon counts with time from Swift/XRT, XMM-Newton, and NuSTAR observations of Mrk 50 at different epochs (see Table 1) with a bin time of 0.2 ks.

**Table A1**
Cross-calibration Factors across All Epochs for All Instruments Used in This Work

| Model | ID | $C_{XRT}$ | $C_{BAT}$ | $C_{OM}$ | $C_{PN}$ | $C_{MOS1}$ | $C_{MOS2}$ | $C_{FPMA}$ | $C_{FPMB}$ |
|---|---|---|---|---|---|---|---|---|---|
| Powerlaw+ | X1 | ⋯ | ⋯ | $1.35^{+0.24}_{-0.26}$ | 1[a] | $0.96 \pm 0.01$ | $0.94 \pm 0.01$ | ⋯ | ⋯ |
| Powerlaw | X2 | ⋯ | ⋯ | ⋯ | 1[a] | $0.94 \pm 0.01$ | $0.92 \pm 0.01$ | ⋯ | ⋯ |
|  | S3+NU | 1[a] | $0.88^{+0.19}_{-0.17}$ | ⋯ | ⋯ | ⋯ | ⋯ | $0.98^{+0.08}_{-0.07}$ | $0.99^{+0.08}_{-0.08}$ |
| Model | ID | $C_{XRT}$ | $C_{BAT}$ | $C_{OM}$ | $C_{PN}$ | $C_{MOS1}$ | $C_{MOS2}$ | $C_{FPMA}$ | $C_{FPMB}$ |
| DiskBB+ | X1 | ⋯ | ⋯ | $2.03^{+0.50}_{-0.41}$ | 1[a] | $0.96 \pm 0.01$ | $0.94 \pm 0.01$ | ⋯ | ⋯ |
| Powerlaw | X2 | ⋯ | ⋯ | ⋯ | 1[a] | $0.94 \pm 0.01$ | $0.93 \pm 0.01$ | ⋯ | ⋯ |
|  | S3+NU | 1[a] | $0.87^{+0.19}_{-0.17}$ | ⋯ | ⋯ | ⋯ | ⋯ | $0.98^{+0.08}_{-0.07}$ | $0.99^{+0.08}_{-0.08}$ |
| Model | ID | $C_{XRT}$ | $C_{BAT}$ | $C_{OM}$ | $C_{PN}$ | $C_{MOS1}$ | $C_{MOS2}$ | $C_{FPMA}$ | $C_{FPMB}$ |
| Optxagnf | X1 | ⋯ | ⋯ | $1.54^{+0.31}_{-0.36}$ | 1[a] | $0.97 \pm 0.01$ | $0.94 \pm 0.01$ | ⋯ | ⋯ |
|  | X2 | ⋯ | ⋯ | ⋯ | 1[a] | $0.94 \pm 0.01$ | $0.93 \pm 0.01$ | ⋯ | ⋯ |
|  | S3+NU | 1[a] | $0.88^{+0.19}_{-0.17}$ | ⋯ | ⋯ | ⋯ | ⋯ | $0.92^{+0.08}_{-0.07}$ | $0.98^{+0.08}_{-0.08}$ |
| Model | ID | $C_{XRT}$ | $C_{BAT}$ | $C_{OM}$ | $C_{PN}$ | $C_{MOS1}$ | $C_{MOS2}$ | $C_{FPMA}$ | $C_{FPMB}$ |
| Powerlaw + Relxillcp | X1 | ⋯ | ⋯ | $1.93^{+0.11}_{-0.17}$ | 1[a] | $0.97 \pm 0.01$ | $0.94 \pm 0.01$ | ⋯ | ⋯ |
|  | X2 | ⋯ | ⋯ | ⋯ | 1[a] | $0.96 \pm 0.01$ | $0.95 \pm 0.01$ | ⋯ | ⋯ |
|  | S3+NU | 1[a] | $0.87^{+0.12}_{-0.18}$ | ⋯ | ⋯ | ⋯ | ⋯ | $0.98^{+0.08}_{-0.07}$ | $0.99^{+0.08}_{-0.08}$ |

**Note.**
[a] Indicates a frozen parameter.

## ORCID iDs

Narendranath Layek ● https://orcid.org/0009-0009-8761-1798
Prantik Nandi ● https://orcid.org/0000-0003-3840-0571
Sachindra Naik ● https://orcid.org/0000-0003-2865-4666
Arghajit Jana ● https://orcid.org/0000-0001-7500-5752

## References

Alexander, T. 1997, in Astrophysics and Space Science Library, Astronomical Time Series, ed. D. Maoz, A. Sternberg, & E. M. Leibowitz, Vol. 218 (Berlin: Springer), 163
Alexander, T. 2013, arXiv:1302.1508
Antonucci, R. 1993, ARA&A, 31, 473
Arnaud, K. A. 1996, in ASP Conf. Ser. 101, Astronomical Data Analysis Software and Systems V, ed. G. H. Jacoby & J. Barnes (San Francisco, CA: ASP), 17
Arnaud, K. A., Branduardi-Raymont, G., Culhane, J. L., et al. 1985, MNRAS, 217, 105
Barth, A. J., Pancoast, A., Thorman, S. J., et al. 2011, ApJL, 743, L4
Barthelmy, S. D., Barbier, L. M., Cummings, J. R., et al. 2005, SSRv, 120, 143
Bennett, C. L., Halpern, M., Hinshaw, G., et al. 2003, ApJS, 148, 1
Bentz, M. C., & Katz, S. 2015, PASP, 127, 67
Bianchi, S., Bonilla, N. F., Guainazzi, M., Matt, G., & Ponti, G. 2009, A&A, 501, 915
Boissay, R., Ricci, C., & Paltani, S. 2016, A&A, 588, A70
Burrows, D. N., Hill, J. E., Nousek, J. A., et al. 2005, SSRv, 120, 165
Cardelli, J. A., Clayton, G. C., & Mathis, J. S. 1989, ApJ, 345, 245
Chakrabarti, S., & Titarchuk, L. G. 1995, ApJ, 455, 623
Chalise, S., Lohfink, A. M., Chauhan, J., et al. 2022, MNRAS, 517, 4788
Crummy, J., Fabian, A. C., Gallo, L., & Ross, R. R. 2006, MNRAS, 365, 1067
Czerny, B., & Elvis, M. 1987, ApJ, 321, 305
Czerny, B., Nikołajuk, M., Różańska, A., et al. 2003, A&A, 412, 317
Dauser, T., Garcia, J., Parker, M. L., Fabian, A. C., & Wilms, J. 2014, MNRAS, 444, L100
Dauser, T., Garcìa, J., Walton, D. J., et al. 2016, A&A, 590, A76
De Marco, B., Ponti, G., Cappi, M., et al. 2013, MNRAS, 431, 2441
den Herder, J. W., Brinkman, A. C., Kahn, S. M., et al. 2001, A&A, 365, L7
Done, C., Davis, S. W., Jin, C., Blaes, O., & Ward, M. 2012, MNRAS, 420, 1848
Done, C., Gierliński, M., & Kubota, A. 2007, A&ARv, 15, 1
Edelson, R., & Malkan, M. 2012, ApJ, 751, 52
Edelson, R., Turner, T. J., Pounds, K., et al. 2002, ApJ, 568, 610
Edelson, R. A., Alexander, T., Crenshaw, D. M., et al. 1996, ApJ, 470, 364
Edelson, R. A., & Krolik, J. H. 1988, ApJ, 333, 646
Ehler, H. J. S., Gonzalez, A. G., & Gallo, L. C. 2018, MNRAS, 478, 4214
Emmanoulopoulos, D., McHardy, I. M., & Papadakis, I. E. 2011, MNRAS, 416, L94
Evans, P. A., Beardmore, A. P., Page, K. L., et al. 2009, MNRAS, 397, 1177
Event Horizon Telescope Collaboration, Akiyama, K., Alberdi, A., et al. 2019, ApJL, 875, L1
Fabian, A. C., Ballantyne, D. R., Merloni, A., et al. 2002, MNRAS, 331, L35
Fabian, A. C., Lohfink, A., Kara, E., et al. 2015, MNRAS, 451, 4375
Fabian, A. C., Zoghbi, A., Ross, R. R., et al. 2009, Natur, 459, 540
Fanali, R., Caccianiga, A., Severgnini, P., et al. 2013, MNRAS, 433, 648
Foreman-Mackey, D. 2017, corner.py: Corner plots, Astrophysics Source Code Library, ascl:1702.002
Gallo, L. C. 2006, MNRAS, 368, 479
Garcìa, J., Dauser, T., Lohfink, A., et al. 2014, ApJ, 782, 76
Garcìa, J., Dauser, T., Reynolds, C. S., et al. 2013, ApJ, 768, 146
Garcìa, J., & Kallman, T. R. 2010, ApJ, 718, 695
Garcìa, J. A., Kara, E., Walton, D., et al. 2019, ApJ, 871, 88
Garcìa, J. A., Steiner, J. F., Grinberg, V., et al. 2018, ApJ, 864, 25
Gehrels, N., Chincarini, G., Giommi, P., et al. 2004, ApJ, 611, 1005
George, I. M., & Fabian, A. C. 1991, MNRAS, 249, 352
Ghosh, R., & Laha, S. 2020, MNRAS, 497, 4213
Ghosh, R., Laha, S., Deshmukh, K., et al. 2022, ApJ, 937, 31
Gierliński, M., & Done, C. 2004, MNRAS, 349, L7
Gliozzi, M., Papadakis, I. E., Grupe, D., Brinkmann, W. P., & Räth, C. 2013, MNRAS, 433, 1709
Goodman, J., & Weare, J. 2010, CAMCS, 5, 65
Haardt, F., & Maraschi, L. 1991, ApJL, 380, L51
Haardt, F., & Maraschi, L. 1993, ApJ, 413, 507
Halpern, J. P. 1984, ApJ, 281, 90
Harrison, F. A., Craig, W. W., Christensen, F. E., et al. 2013, ApJ, 770, 103
Jana, A., Kumari, N., Nandi, P., et al. 2021, MNRAS, 507, 687
Jansen, F., Lumb, D., Altieri, B., et al. 2001, A&A, 365, L1
Jiang, J., Parker, M. L., Fabian, A. C., et al. 2018, MNRAS, 477, 3711
Kormendy, J., & Richstone, D. 1995, ARA&A, 33, 581
Krolik, J. H. 1999, in Active Galactic Nuclei : from the Central Black Hole to the Galactic Environment, Vol. 420 (Princeton, NJ: Princeton Univ. Press), L57
Kubota, A., & Done, C. 2018, MNRAS, 480, 1247
Kumari, N., Jana, A., Naik, S., & Nandi, P. 2023, MNRAS, 521, 5440
Laha, S., Guainazzi, M., Dewangan, G. C., Chakravorty, S., & Kembhavi, A. K. 2014, MNRAS, 441, 2613
Laor, A. 1991, ApJ, 376, 90
Layek, N., Nandi, P., Naik, S., et al. 2024, MNRAS, 528, 5269
Leighly, K. M. 1999a, ApJS, 125, 297
Leighly, K. M. 1999b, ApJS, 125, 317
Madathil-Pottayil, A., Walton, D. J., Garcìa, J., et al. 2024, MNRAS, 534, 608
Magdziarz, P., Blaes, O. M., Zdziarski, A. A., Johnson, W. N., & Smith, D. A. 1998, MNRAS, 301, 179
Mason, K. O., Breeveld, A., Much, R., et al. 2001, A&A, 365, L36
Matt, G., Perola, G. C., & Piro, L. 1991, A&A, 247, 25
Matzeu, G. A., Nardini, E., Parker, M. L., et al. 2020, MNRAS, 497, 2352
Middei, R., Tombesi, F., Vagnetti, F., et al. 2020, A&A, 635, A18
Middleton, M., Done, C., Ward, M., Gierliński, M., & Schurch, N. 2009, MNRAS, 394, 250
Miniutti, G., & Fabian, A. C. 2004, MNRAS, 349, 1435
Miniutti, G., Ponti, G., Greene, J. E., et al. 2009, MNRAS, 394, 443
Mitsuda, K., Inoue, H., Koyama, K., et al. 1984, PASJ, 36, 741
Nandi, P., Chatterjee, A., Chakrabarti, S. K., & Dutta, B. G. 2021, MNRAS, 506, 3111
Nandi, P., Chatterjee, A., Jana, A., et al. 2023, ApJS, 269, 15
Nandi, P., Naik, S., Chatterjee, A., et al. 2024, MNRAS, 532, 1185
Nandra, K., George, I. M., Mushotzky, R. F., Turner, T. J., & Yaqoob, T. 1997, ApJ, 476, 70
Nandra, K., & Papadakis, I. E. 2001, ApJ, 554, 710
Narayan, R., & Yi, I. 1994, ApJL, 428, L13
Netzer, H. 2013, in The Physics and Evolution of Active Galactic Nuclei (Cambridge: Cambridge Univ. Press)
Noda, H., & Done, C. 2018, MNRAS, 480, 3898
Pal, M., Dewangan, G. C., Kembhavi, A. K., Misra, R., & Naik, S. 2018, MNRAS, 473, 3584
Pancoast, A., Brewer, B. J., Treu, T., et al. 2012, ApJ, 754, 49
Pastoriza, M. G., Bica, E., Bonatto, C., Mediavilla, E., & Perez, E. 1991, AJ, 102, 1696
Peterson, B. M., Ferrarese, L., Gilbert, K. M., et al. 2004, ApJ, 613, 682
Petrucci, P. O., Gronkiewicz, D., Rozanska, A., et al. 2020, A&A, 634, A85
Petrucci, P. O., Ursini, F., De Rosa, A., et al. 2018, A&A, 611, A59
Piconcelli, E., Jimenez-Bailón, E., Guainazzi, M., et al. 2005, A&A, 432, 15
Porquet, D., Reeves, J. N., Hagen, S., et al. 2024, A&A, 689, A336
Porquet, D., Reeves, J. N., O'Brien, P., & Brinkmann, W. 2004, A&A, 422, 85
Pounds, K. A., Warwick, R. S., Culhane, J. L., & de Korte, P. A. J. 1986, MNRAS, 218, 685
Rees, M. J. 1984, ARA&A, 22, 471
Ricci, C., Ho, L. C., Fabian, A. C., et al. 2018, MNRAS, 480, 1819
Rodrìguez-Pascual, P. M., Alloin, D., Clavel, J., et al. 1997, ApJS, 110, 9
Roming, P. W. A., Kennedy, T. E., Mason, K. O., et al. 2005, SSRv, 120, 95
Ross, R. R., & Fabian, A. C. 1993, MNRAS, 261, 74
Ross, R. R., & Fabian, A. C. 2005, MNRAS, 358, 211
Schlafly, E. F., & Finkbeiner, D. P. 2011, ApJ, 737, 103
Shakura, N. I., & Sunyaev, R. A. 1973, A&A, 24, 337
Singh, K. P., Garmire, G. P., & Nousek, J. 1985, ApJ, 297, 633
Strateva, I. V., Brandt, W. N., Schneider, D. P., Vanden Berk, D. G., & Vignali, C. 2005, AJ, 130, 387
Strüder, L., Briel, U., Dennerl, K., et al. 2001, A&A, 365, L18
Sun, W.-H., & Malkan, M. A. 1989, ApJ, 346, 68
Sunyaev, R. A., & Titarchuk, L. G. 1980, A&A, 86, 121
Tananbaum, H., Avni, Y., Branduardi, G., et al. 1979, ApJL, 234, L9
Tripathi, P., & Dewangan, G. C. 2022, ApJ, 925, 101
Tripathi, P., Dewangan, G. C., Papadakis, I. E., & Singh, K. P. 2021, ApJ, 915, 25
Tripathi, S., Waddell, S. G. H., Gallo, L. C., Welsh, W. F., & Chiang, C. Y. 2019, MNRAS, 488, 4831
Turner, M. J. L., Abbey, A., Arnaud, M., et al. 2001, A&A, 365, L27
Turner, T. J., & Pounds, K. A. 1989, MNRAS, 240, 833
Turner, T. J., Reeves, J. N., Braito, V., & Costa, M. 2018, MNRAS, 476, 1258

Ursini, F., Petrucci, P. O., Bianchi, S., et al. 2020, A&A, 634, A92
Vasudevan, R. V., Brandt, W. N., Mushotzky, R. F., et al. 2013, ApJ, 763, 111
Vasudevan, R. V., & Fabian, A. C. 2007, MNRAS, 381, 1235
Vaughan, S., Edelson, R., Warwick, R. S., & Uttley, P. 2003, MNRAS, 345, 1271
Vaughan, S., Fabian, A. C., Ballantyne, D. R., et al. 2004, MNRAS, 351, 193
Waddell, S. G. H., Gallo, L. C., Gonzalez, A. G., Tripathi, S., & Zoghbi, A. 2019, MNRAS, 489, 5398
Walter, R., & Fink, H. H. 1993, A&A, 274, 105
Walton, D. J., Nardini, E., Fabian, A. C., Gallo, L. C., & Reis, R. C. 2013, MNRAS, 428, 2901
Walton, D. J., Reis, R. C., & Fabian, A. C. 2010, MNRAS, 408, 601
Wilms, J., Allen, A., & McCray, R. 2000, ApJ, 542, 914
Xu, Y., Garcìa, J. A., Walton, D. J., et al. 2021, ApJ, 913, 13
Yu, Z., Jiang, J., Bambi, C., et al. 2023, MNRAS, 522, 5456
Zdziarski, A. A., Johnson, W. N., & Magdziarz, P. 1996, MNRAS, 283, 193
Życki, P. T., Done, C., & Smith, D. A. 1999, MNRAS, 309, 561